# Derivatives Holdings and Systemic Risk in the U.S. Banking Sector

María Rodríguez-Moreno
Universidad Carlos II de Madrid
Department of Business Administration
C/ Madrid, 126
28903 Getafe (Madrid, Spain)
mrodri1@emp.uc3m.es

Sergio Mayordomo[1]
University of Navarra
School of Economics and Business Administration
Edificio Amigos
31009 Pamplona (Spain)
Email: smayordomo@unav.es
Fax: + 34 948 425626

Juan Ignacio Peña
Universidad Carlos III de Madrid
Department of Business Administration
C/ Madrid, 126
28903 Getafe (Madrid, Spain)
ypenya@eco.uc3m.es

[1] Corresponding author: Maria Rodriguez-Moreno. The authors acknowledge Germán López-Espinosa, Óscar Arce, Chen Zhou, Simone Manganelli, Stefano Corradin, David Marques-Ibáñez and seminar participants at European Central Bank, FDIC, 29th GdRE Annual International Symposium on Money, Banking and Finance, XX Edition of the Finance Forum AEFIN, CNMV, Universidad Autónoma de Madrid, University of Navarra, for their useful comments and suggestions. The opinions in this article are the sole responsibility of the authors.

# Derivatives Holdings and Systemic Risk in the U.S. Banking Sector

**Abstract**

This paper studies the impact of the banks' portfolio holdings of financial derivatives on the banks' individual contribution to systemic risk over and above the effect of variables related to size, interconnectedness, substitutability, and other balance sheet information. Using a sample of 95 U.S. bank holding companies from 2002 to 2011, we compare five measures of the banks' contribution to systemic risk and find that the new measure proposed in this study, Net Shapley Value, outperforms the others. Using this measure we find that the banks' holdings of foreign exchange and credit derivatives increase the banks contributions to systemic risk whereas holdings of interest rate derivatives decrease it. Nevertheless, the proportion of non-performing loans over total loans and the leverage ratio have much stronger impact on systemic risk than derivatives holdings. Therefore, the derivatives' impact plays a second fiddle in comparison with traditional banking activities related to the former two items.

## 1. Introduction

Since the beginning of the current financial and economic crisis, the concern about systemic risk has increased, becoming a priority for regulatory authorities. The International Monetary Fund, Bank of International Settlements and Financial Stability Board define systemic risk as the "risk of disruption to financial services that (i) is caused by an impairment of all or parts of the financial system and (ii) has the potential to have serious negative consequences for the real economy". Regulatory authorities realized that systemic risk is not a transitory problem and consequently, new institutional arrangements have been approved to address this challenging issue. The Financial Stability Oversight Council (FSOC) in the U.S. and the European Systemic Risk Board (ESRB) in the E.U. have been set to identify systemic risk, prevent regulatory loopholes, and make recommendations together with existing regulatory authorities. The concerns about systemic risk have also extended to securities markets regulators. Thus, the International Organization of Securities Commissions' (IOSCO) has also established a Standing Committee on Risk and Research to coordinate members' monitoring of potential systemic risks within securities markets.

In this setting it is crucial for the banking regulatory institutions to be able to analyze and understand the determinants of a banks' contribution to systemic risk. This information would help them not only to improve currently available systemic risk measures and warning flags but also to develop a taxation system on the basis of the externalities generated by a banks' impact on systemic risk. Additionally, securities market regulators are interested in understanding the contribution of traded financial instruments, for instance financial derivatives, to systemic risk in order to consider new regulatory initiatives. Finally, investors should be concerned with the extent to which derivatives holdings affect the systemic impact of a given bank in order to assess the appropriate reward required to bear this kind of risk. Stulz (2010) pointed out the lack of rigorous empirical studies on the social benefits and costs of derivatives and in particular their role in the financial crisis 2007-09. This paper aims to improve our understanding of these social costs and benefits examining whether the use of financial derivatives was a relevant factor in the destabilization of the banking system during the recent financial crisis.

The spectacular growth in banks' balance sheet over recent decades reflected increasing claims within the financial system rather than with non-financial agents. One key driver of this explosive intra-system activity came from the growth in derivatives markets and consequently in the growth of derivatives holdings in the banks' balance-sheets. A proportion of this growth may have been motivated by their use for hedging purposes justified by theory supporting the rationality of hedging decisions at individual bank level (e.g., Koppenhaver, 1985). This stance also finds support in empirical evidence suggesting the advantages of different hedging strategies for financial firms, again at individual level, see among others Jaffe (2003). However, another substantial proportion of this growth is due to proprietary trading activities by banks. Both activities, hedging and trading, are regarded as potentially useful and profitable by banks. However, it is well known that financial decisions that are rational at individual level can have negative consequences at system level. Is this also the case with respect to the banks' holdings of financial derivatives? The, admittedly very scarce, literature on this subject suggests that this might be the case, Calmès and Théoret (2010) find that off-balance sheet activities reduce banks' mean returns, simultaneously increasing the volatility of their operating revenue and therefore increasing banks' systemic risk. Nijskens and Wagner (2011) report that the first use of credit derivatives is associated with an increase in a bank's risk, largely due to an increase in banks' correlations and therefore in their systemic risk. However, as far as we know, no evidence is available on the direct impact of derivatives holdings on the banks' individual contributions to systemic risk. Ours is a first attempt to fill this gap. For such aim, we combine two analyses; we first measure the banks' individual contributions to systemic risk and then, we estimate the effects of their holdings of financial derivatives on the banks' contributions to systemic risk.

To assess the banks' contributions to systemic risk we use the following five measures: ΔCoVaR, ΔCoES, Asymmetric ΔCoVaR, Gross Shapley Value (GSV) and Net Shapley Value (NSV). The ΔCoVaR is the difference between the Value at Risk (VaR) of the banking system conditional on bank $i$ being in distress minus the VaR of the banking system conditional on bank $i$ being in its median state. The ΔCoES applies the same idea but using the Expected Shortfall instead of the VaR (see Adrian and Brunnermeier, 2011). The Asymmetric ΔCoVaR represents a variation of the standard ΔCoVaR specification that allows for asymmetries in this specification (see Lopez, Moreno, Rubia and Valderama, 2011). The GSV measures the average contribution to systemic risk of bank $i$ in all possible groups in which the whole financial system can be divided (see Tarashev, Borio, and Tsatsaronis, 2010). Finally we propose an alternative measure to the GSV called NSV in which we get rid of the idiosyncratic component present in the former measure by subtracting from the GSV the VaR of the bank $i$.

We estimate these five measures for a subset of the 95 biggest U.S. bank holding companies for the period that spans from 2002 to 2011. We then compute the correlation of the systemic risk measures with an index of systemic events and run a Granger causality test between pairs of measures; and find that the NSV presents the closest association with the index and Granger causes more frequently the other measures. Then, using this measure of systemic risk as the dependent variable, we analyze the effect of the banks' portfolio holdings of financial derivatives on the banks' individual contribution to systemic risk over and above the effect of variables related to size, interconnectedness, substitutability, and other balance sheet information.

The main results of the paper can be summarized as follows. We find a significant relationship between the fair values of derivatives holdings of given bank in a given quarter and the bank's contribution to systemic risk one quarter later. Therefore, derivatives holdings act as leading indicators of systemic risk contributions. Nevertheless, this relationship is not uniform across derivatives classes. Banks' holdings of credit and foreign exchange derivatives have an increasing effect on systemic risk whereas holdings of interest rate derivatives have a decreasing effect.

Given that the baseline model directly relates the banks' derivatives activity to their individual contribution to systemic risk, the results may potentially suffer for endogeneity problems in the sense that banks with certain specific characteristics self-select them into derivatives participation. To address this concern we provide four pieces of evidence that enable us to confirm that our findings come indeed from a causal relation between banks' derivatives holdings and their contribution to systemic risk.

Besides derivatives there are other balance sheet items that are significant contributors to systemic risk. Thus, increases in the following variables (measured as ratios over total assets) led to increases in systemic risk contributions: total loans, net balance to banks belonging to the same banking group, leverage ratio, and the proportion of non-performing loans (measured in this case, relative to total loans). On the other hand, increases in total deposits decrease systemic risk. The variables with the highest economic impact on systemic risk are the proportion of non-performing loans to total loans and the leverage ratio. In fact, their economic impact is much higher than the one corresponding to derivatives holdings. Therefore, the derivatives' impact plays a second fiddle in comparison with traditional banking activities related to the former two items.

The rest of the paper is organized as follows. Section 2 describes the methodology. In Section 3 we describe the data. Section 4 reports the main empirical findings. In Section 5 we present some robustness tests, and we conclude in Section 6.

## 2. Methodology

### 2.1. Systemic Risk: Measures and Comparison

We consider the following five measures of the individual contribution of banks to systemic risk: (i) ΔCoVaR, (ii) ΔCoES, (iii) Asymmetric ΔCoVaR, (iv) Gross Shapley Value (GSV) and (v) Net Shapley Value (NSV). The details of the characteristics and the estimation of the systemic risk measures can be found in Appendix B.[2]

As in Rodriguez-Moreno and Peña (2013) we use two criteria to rank the five measures: (a) the correlation with an index of systemic events and policy actions, and (b) the Granger causality test. The first criterion compares the correlation of each measure with the main systemic events and policy actions and the second criterion points out the measures acting as leading indicators of systemic risk. Both criteria focus on different aspects of systemic risk and complement to each other to provide a robust diagnostic of the most reliable individual contribution to systemic risk measures.[3]

In the first criterion we use an influential event variable (IEV), which is a categorical variable that captures the main events observed and policy actions taken during the financial crisis based on the Federal Reserve Bank of St. Louis' crisis timeline.[4] The IEV takes value 1 whenever there is an event, under the hypothesis that those events should increase systemic risk, and is equal to -1 whenever there is a policy action, under the hypothesis that policy action's aim is to decrease systemic risk. Otherwise it equals zero. The previous influential event variable lies on the assumption that policy actions succeed in diminishing systemic risk. Given that the assumption that policy actions are totally effective could seem strong, we use an alternative influential event variable to compare the measures that ignores policy actions (i.e. it assigns a value of 0 to them instead of -1).

---

[2] Acharya, Pedersen, Philippon and Richardson, (2011a, b) propose an alternative measure of the individual contribution to systemic risk called realized SES that measures the propensity of bank $i$ to be undercapitalized when the whole system is undercapitalized. We exclude this measure from the discussion in the main text because, by construction, it is quarterly estimated and we cannot carry out the comparison with the considered five measures. Nevertheless, we estimate this measure, conduct the baseline regression to analyze the determinants of banks contributions to systemic risk and find that the results are fully in agreement with the main findings of this paper.
[3] In Rodriguez-Moreno and Peña (2013) the authors use an additional criterion based on the Gonzalo and Granger's (1995) methodology. To carry out this analysis, the pairs of systemic risk measures have to be cointegrated. However, this requirement is not satisfied in several of the pairs of measures and so, we do not consider it.
[4] Timeline crisis can be accessed via http://timeline.stlouisfed.org/.

The ranking method for the two variations of the first criterion is based on the McFadden R-squared, a measure of goodness of fit. For each bank $i$ in the sample we run a regression in which the dependent variable is the IEV and the explanatory variable is the systemic risk measure $j$ for bank $i$ (where $j = 1, \ldots, 5$ and $i = 1, \ldots, 95$) and then estimate the McFadden R-squared. When the influential event variable includes the policy actions we run a multinomial logistic regression but when it does not we run a logistic regression. The comparison of the different pairs of systemic risk measures, referred to the same bank, is done by assigning a score of +1 to the measure with the highest R-squared and -1 to the one with the lowest. Finally, we add up the scores obtained for each measure across the 95 banks.[5] By doing this, we avoid penalizing those measures that provide leading information and penalizing those events or political actions which have been discounted by the market before the event.

The second criterion is based on the Granger causality test (Granger, 1969). To rank the measures we give a score of +1 to a given measure X if X Granger causes another measure Y at 5% confidence level and -1 if X is caused in the Granger sense by Y. As a consequence, the best measure gets the highest positive score and the worst measure the highest negative score. Next, we add up the scores obtained by each measure across the 95 banks. Technical details on the procedure to compare the systemic risk measures can be found in Appendix C.

**2.2. Determinants of systemic risk**

We implement a panel regression analysis in which the individual bank $i$'s contribution to systemic risk in quarter $t$ is regressed on the following variables (all in quarter $t-1$): bank's holdings of derivatives, proxies for the standard drivers of systemic risk (size, interconnectedness, and substitutability), other balance sheet information and the aggregate level of systemic risk. In our baseline analysis we employ a Prais-Winsten regression with correlated panels, corrected standard errors (PCSEs) and robust to heteroskedasticity and contemporaneous correlation across panels. Our panel regression model is described by the following equation:

$$SR_{i,t} = \alpha + \sum_{n=1}^{N} \gamma_n Y_{n,i,t-1} + \sum_{m=1}^{M} \omega_m Z_{m,i,t-1} + \sum_{s=1}^{S} \beta_s X_{s,i,t-1} + \varepsilon_{i,t} \qquad (1)$$

where the dependent variable is the bank's $i$ contribution to systemic risk as measured by the Net Shapley Value. The vector of variables $Y_{n,i,t}$ contains the proxies for the bank $i$ size and its degree of interconnectedness and substitutability. The vector $Z_{m,i,t}$ contains variables related to other banks characteristics: balance sheet quality and the aggregate level of systemic risk one and two quarters ago. The aggregate variables are obtained after aggregating the levels of systemic risk of the U.S. commercial banks (without considering the bank $i$), dealer-broker and insurance companies. The vector of variables $X_{s,i,t}$ refers to the banks' holdings of financial derivatives.

---

[5] This ranking procedure is related to the well-known Condorcet voting method. However to avoid some of the problems of the Condorcet approach we also allow for negative as well as positive scores.

### 2.3. Research questions

We examine three issues that have not been addressed previously in literature regarding the role of derivatives holdings and their possible connections with systemic risk:

1. The first question to ask is whether the banks' holdings of financial derivatives contribute in any significant way to systemic risk.
2. The next obvious question is whether this relationship is uniform across derivatives classes or are there differences in the impact between foreign exchange and interest rate derivatives, for example.
3. Additionally, it seems natural to ask what other balance sheet asset items are significant contributors to systemic risk and in particular which ones have the biggest economic impact on systemic risk.

## 3. Data and Explanatory Variables

### 3.1. Data

The Bank Holding Company Data (BHCD) from the Federal Reserve Bank of Chicago is our primary database. Additional information (VIX, 3-monthTbill rate, 3-month repo rate, 10-year Treasury rate, BAA-rate bond, and MSCI index returns) is collected from DataStream and the Federal Reserve Bank of New York.

Our data set is composed of U.S. bank holding companies with total assets above $5billion in either the first quarter of 2006 or the first quarter of 2009. This criterion allows us to cover a representative sample of U.S bank holdings companies (our sample covers, e.g., on average, we cover the 70% of all U.S. bank holding companies' total asset value) considering both systemically important financial institution (SIFIs) and non-SIFIs.[6,7] This criterion is applied during the pre-crisis and the ongoing crisis period in order to avoid any potential bias. Additionally, we remove those banks for which we do not have information on their stock prices and, banks that defaulted or were acquired before the end of 2006.[8] Our final sample consists of quarterly information for 95 bank holding companies from March 2002 to June 2011.

Table I contains the 95 banks and information about their size (average market capitalization in millions of dollars during the sample period). In terms of size we observe a huge variance across banks under analysis with Bank of America, Citigroup and JP Morgan being the largest banks in the sample by a wide margin.

**3.2. Explanatory Variables**

Next we summarize the five groups of potential determinants of the banks' contribution to systemic risk (a detailed description can be found in Appendix A).

### 3.2.1. Banks Holdings of Derivatives

We consider five types of derivatives: credit, interest rate, foreign exchange, equity, and commodity. Although previous literature about the effect of financial derivatives on systemic risk is scarce, some papers suggest the possible role of credit derivatives as determinant of systemic risk (see Stulz, 2004 and Acharya, 2011). Moreover, the hedging offered by derivatives could also lead banks to take more risk on the underlying asset. This fact could destabilize the banking sector if markets are not perfectly competitive (Instefjord, 2005).

---

[6] The implementation of enhanced prudential standards in the Dodd-Frank Act generally applies to all U.S. bank holding companies with consolidated assets of $50 billion or more.
7 In their analysis, Acharya et al. (2011b) use financial institutions whose market cap exceeds $5billion as of end of June 2007. We use a less restrictive criterion to include more banks in our analysis and apply the same threshold ($5billion) but over total assets. Thus, by June 2007 the market cap of the smallest bank in our sample was around $3.3 billion.
[8] We deal with bank mergers as in Hirtle (2008) who adjusts for the impact of significant mergers by treating the post-merger bank as a different entity from the pre-merger bank. This is the case of the Bank of New York Company and Mellon Financial Corp.

The holdings of derivatives are considered in terms of the fair value that is defined in the instructions of preparation of the BHCD as "the price that would be received to sell an asset or paid to transfer a liability in an orderly transaction between market participants in the asset's or liability's principal (or most advantageous) market at the measurement date". The holdings of derivatives are reported in the balance sheet with positive (asset side) or negative (liabilities side) fair values which refer to the amount of revaluation gains or losses from the ''marking to market'' of the five different types of derivative contracts.[9, 10] We focus on the total fair value (i.e., positive plus negative fair values) because it allows us to take into account the total exposures to the derivatives' counterparties and, at the same time, the counterparty risk. Alternatively to the fair value, we could use the notional amount outstanding. According to the Office of the Comptroller of the Currency (OCC) Quarterly Reports on Bank Trading and Derivatives Activities notional values can provide insight into potential revenue and operational issues but do not provide useful measure of the risk taken and so, could be meaningless from the systemic risk perspective.[11] Nevertheless, later robustness checks show that results do not depend on the use of the fair value or notional amount to value the banks' holdings of credit derivatives.

### 3.2.2. Size

The impact of size on systemic risk is increasing and possibly non-linear as documented in Pais and Stork (2011). Tarashev, Borio and Tsatsaronis (2010) convincingly argue that larger size implies greater systemic importance, that the contribution to system-wide risk increases more than proportionately with relative size, and that a positive relationship between size and systemic importance is a robust result. The logarithm of the market capitalization (share price multiplied by the number of ordinary shares in issue) is used as the proxy for its size. This is a common practice in finance (e.g. Ferreira and Laux, 2007) and accounting (e.g. Bhen, Choi, and Kang, 2008) literature. We use market value instead of total assets to avoid any collinearity problem because banks' total assets have been employed to define and standardize most of the variables. We add the square of the size variable to our regression to control any potential non-linear relation between size and systemic risk.

---

[9] Unlike other securities, derivative contracts involve two possible positions and positive fair values mean negative fair values on the counterparty. According to the Dodd-Frank Act, the required information to private funds advised by investment advisers to guarantee an appropriate monitoring of systemic risk in securities markets includes: amount of assets under management and use of leverage, trading and investment positions, types of assets held, or trading practices, among others contracts.

[10] The statement of Financial Accounting Standard No. 133 "Accounting for Derivative Instruments and Hedging Activities" requires all derivatives, without exception and regardless of the accounting treatment of the underlying asset, to be recognized in the balance sheet as either negative fair values (liabilities) or positive fair values (assets).

[11] There are several examples of studies in previous literature using the derivatives fair value (e.g. Venkatachalam, 1996; or Livne, Markarian and Milne, 2011, among others).

### 3.2.3. Interconnectedness and substitutability

Interconnectedness measures the extent to which a bank is connected with other institutions in such a way that its stress could easily be transmitted to other institutions. Substitutability can be defined as the extent to which other institutions or segments of the financial system can provide the same services that were provided by failed institutions. These two concepts are not easy to measure and there is therefore scarce evidence quantifying their effects on systemic risk.

As pointed out by Acharya, Pedersen, Philippon, and Richardson (2011a), the dimensions of systemic risk can be also translated into the following groups: size, leverage, risk, and correlation with the rest of the financial sector and economy. Due to the difficulty of measuring substitutability and interconnectedness, they are grouped in a more general group: correlation of the bank with the financial sector and economy.

To control for these dimensions we first employ some variables that could be more related to the interconnectedness dimension and then other variables related to the substitutability dimension. In the first group we consider the net balances to subsidiary banks and non-banks as a way to study the net position of a bank within the group. Additionally, this first dimension is captured by means of the correlation between the average daily individual bank's stock returns and the S&P500 index returns during the corresponding calendar quarter $t$ (hereafter correlation with S&P500 index) in line with Allen, Bali, and Tang (2012).

In the second group we include variables related with the substitutability as reflected into the services that are provided by the banks, and we also distinguish between variables referred to the core and non-core banking activities. Brunnermeier, Dong and Palia (2011) find that non-interest to interest income variable (proxy for the non-core or non-traditional activities such as trading and securitization, investment banking, brokerage or advisory activities) has a significant contribution to systemic risk; we include this variable in our regressions. On the other hand, the amount of loans to banks and depository institutions relative to total assets and the total loans (excluding loans to banks and depository institutions) relating to total assets represent the bank's core or traditional activities. We distinguish between loans to the financial system and other loans enabling us to study whether they have different effects on systemic risk. Finally, we use the ratio of the bank's commercial paper holding relative to total assets as a proxy for the interbank activities given that we do not have direct information on the interbank lending. As Cummins and Weiss (2010) state, the inter-bank lending and commercial paper markets were critical in the subprime crisis. These variables could also indicate to some extent the degree of interconnectedness of a given bank given that the larger the total amount of the loans the larger is the expositions of a given bank to their borrowers. The difficulty of defining proxies related to the bank degree of substitutability could be one of the reasons that explain the scarcity of studies quantifying the effect of this dimension of systemic risk.[12] We define the variables referred to interconnectedness relative to the bank total assets.

### 3.2.4. Balance Sheet Information

We use several variables that refer to the balance sheet quality: (i) leverage, (ii) total deposits relative to total assets, (iii) maturity mismatch, and (iv) non-performing loans to total loans.

One of the dimensions proposed by Acharya, Pedersen, Philippon, and Richardson (2011b) is leverage, however true leverage is not straightforward to measure due to the limited market data breaking down off- and on-balance sheet financing. According to them we define leverage as follows:

$$Leverage = \frac{book\ assets - book\ equity + market\ equity}{market\ value\ of\ equity} \qquad (2)$$

---

[12] We are aware of only one study analyzing the effect of the substitutability dimension on systemic risk: Cummings and Weiss (2010). The authors study whether the U.S. insurers' activities create systemic risk and show that the lack of substitutability of insurers is not a serious problem. According to their results even a default of large insurers would not create a substitutability problem because other insurers could fill this gap. However, we consider that banking sector differs from the previous one and for this reason a positive effect of the substitutability dimension on the bank contribution to systemic risk cannot be ruled out.

As pointed out by Acharya and Thakor (2011) higher bank leverage creates stronger creditor discipline at individual bank level but it also increases systemic risk. However, some empirical analyses do not find significant effect of leverage on systemic risk (see Brunnermeier et al., 2011; or Lopez, et al., 2011). Mizrach (2011) shows conventionally measured leverage as an unreliable indicator of systemic risk and suggests a more detailed examination of bank balance sheets and asset holdings.

Two other potential explanatory variables are maturity mismatch and deposits to total assets. Thus, the higher the mismatch the more likely the bank is exposed to funding stress. Deposits to total assets have two different interpretations. On the one hand during financial distress periods banks could rely more on deposits (see Boyson, Helwege, and Jindra, 2011). On the other hand, activities that are not traditionally associated with banks (outside the realm of traditional deposit taking and lending) are associated with a larger contribution to systemic risk and activities related to deposits taking are associated with a lower contribution to systemic risk. Total deposits could contribute to decrease systemic risk because they provide a shock-absorbing buffer.

Regarding the ratio of non-performing loans to total loans, the growth of credit and the easy access to financing observed before the subprime crisis (Dell'Ariccia et al., 2008) could have increased substantially the role of this variable as a significant determinant of the bank's contribution to systemic risk.

### 3.2.5. Aggregate systemic risk measure

The aggregate systemic risk for each bank $i$ is estimated as the sum of the individual contribution to systemic risk of all the banks with the exception of bank $i$, the 8 major broker-dealers, and the 23 major insurance companies.[13] This variable captures the deterioration of the financial system's health. We use two lags of the aggregate measure of systemic risk to control by speed of adjustment to the aggregate level of risk and to absorb any lagged aggregated information transmitted into the current observation.

Table II reports the main descriptive statistics of the explanatory variables in the baseline analysis. We observe that the holdings of financial derivatives represent, on average, a small proportion of the total assets. They range from the interest rate derivatives, averaging 3% of total assets to commodity derivatives averaging only 0.1%. Interest rate derivatives are the more frequent type of derivative in the banks' balance sheets: only 4 out of 95 banks forming the sample did not hold interest rate derivatives at any quarter of the sample period. The other types of derivatives are less frequent in the banks' balance sheets: 35, 65, 69, and 73 banks never held foreign exchange, credit, equity, and commodity derivatives, respectively. Net balances due to bank represent, on average, a lower proportion than net balances due to non-banks. The average correlation of the individual banks with S&P500 index is quite large (0.59) which suggests a substantial interconnectedness of the banking system with the overall market. Average total loan and loan to banks represent around 61% and 0.2% of the total assets, respectively. The average ratio non-interest to interest income is close to 0.5 and average maturity mismatch is close to 10%. Finally, the balance sheet category, total deposits represent, on average, almost 70% of total assets.[14]

---

[13] The sample of the 8 major broker-dealers and the 23 major insurance companies corresponds to the institutions studied in Acharya et al. (2011b) for which we have information on their stock prices. We remove those institutions that defaulted or were acquired before the end of 2006.

Average correlations among the explanatory variables are reported in Table III. The absolute values of the correlations among explanatory variables are rarely higher than 0.5 when we do not consider correlations among holdings of derivatives related variables. Nevertheless, there are a few exceptions that are worth mentioning. We observe that bank size is positively correlated with the commercial paper activity and negatively correlated with the deposits activity. This is consistent with larger banks being less involved in traditional activities. As expected, the non-interest to interest income is negatively correlated with the loans. The non-performing loans are positively correlated with the leverage, suggesting that the most leveraged banks are the ones with higher delinquencies ratios.

Looking at the correlation across derivatives holdings, we observe that all of them are strongly interconnected suggesting that banks that use a given type of derivative tend to use the other derivatives. The correlations among all the types of derivatives holdings with the bank's size, the commercial paper, the loans to banks, the non-interest to interest income, the correlation with the S&P500 returns, the leverage, the maturity mismatch, and the non-performing loans; are always positive. On the other hand the correlations with the deposits are negative for all the derivatives holdings.

---

[14] We consider more relevant the use of an aggregate lagged systemic risk rather than the lags of the dependent variable. Nevertheless, we repeat the analysis excluding the aggregate systemic risk measure and including time effects and two lags of the bank contribution to systemic risk and obtain similar results.

Overall, the signs of the correlations among the explanatory variables are in agreement with the expected ones. Additionally, the magnitudes of the correlations are not high enough to indicate potential problems of collinearity. To check this statement we compute the scaled condition index for the variables in Table III and find that it is below the threshold compatible with the absence of collinearity (30). Our main objective, however, is to examine the joint effect of these explanatory variables, highlighting the role of the banks' portfolio holding of the indicator of the individual contribution to systemic risk by means of a regression analysis.

## 4. Empirical Results

### 4.1. Individual Systemic Risk Measures and Their Comparison

Panel A of Table IV reports the main descriptive statistics of the individual quarterly measures. The signs for all the measures are set such that the higher the measure, the higher the bank's contribution to systemic risk. The measures are defined in basis points. We observe a common pattern in all of them with a huge difference between the mean and the maximum due to the big jump during Lehman Brothers episode.

We then rank the systemic risk measures according to the two criteria stated in Section 2.1 and Appendix C: (a) the correlation with an index of systemic events and policy actions and (b) Granger causality test. Panels B and C of Table IV contain the final scores. Panel B corresponds to the case in which the influential event variable considers the effect of the policy actions and assigns them a value of -1 while Panel C corresponds to the case in which no policy actions are included among the influential events and so the events variable takes value 1 if there is a systemic event and 0 otherwise. Comparing the five weekly measures, we observe in Panels B and C that under both criteria, the NSV obtains the highest score followed by the GSV. Additionally, we compare the five systemic risk measures for a portfolio that consists of only the 16 largest banks and confirm that the NSV is the most reliable measure.[15] Therefore, for the baseline analysis we use the NSV as the proxy for the bank contribution to systemic risk. Some robustness checks using alternative measures of systemic risk are conducted in Section 5.

---

[15] In fact, the pairwise correlation between the NSV estimated in the baseline analysis and the NSV using the portfolio of the largest 16 banks is, on average, 0.99 for the group formed by these 16 banks.

Additional aspects of the different measures are worth mentioning. The co-risk measures strongly rely on the performance of the state variables and employ little firm specific information (i.e., information contained on stock prices, total assets and book equity). So, these measures provide very similar output for different banks independent of the bank's risk profile. To give an example, the estimation of CoVaR for every bank $i$ (Equations B.1.1-B.1.3) is done using the growth rate of the market value of total financial assets (at system level) as the dependent variable; and a set of state variables and the growth rate of the market value of total financial assets of bank $i$ as explanatory variables. The results of the quantile regression shows that the coefficient measuring the impact of the market value of the total financial assets of bank $i$ on this measure of systemic risk is significant only for 11 of the 95 banks at 10% of significance level when quantile level is 1% ($q = 0.01$) and in zero cases when quantile level is 50% ($q = 0.5$). Therefore individual bank's CoVaR is largely determined by the same set of common variables. For this reason, we expect strong similarities across banks in terms of this systemic risk measure.[16]

Regarding the computation of the GSV for bank $i$, this measure includes the VaR of bank $i$ as an additional element in estimating the individual contribution to systemic risk. But in non-stress periods (where the individual contribution of bank $i$ to system risk is negligible) this measure is largely determined by the evolution of the VaR of bank $i$ which is a measure of the bank's individual risk.[17] Thus, we consider an alternative measure which is net of the impact of a proportion of the individual VaR, the Net Shapley Value. That is, we get rid of the bank's idiosyncratic risk and focus on the bank's contribution to systemic risk by subtracting the VaR from the GSV. On the other hand, one may also argue that the NSV could be a poorer measure of the externality that the risk taking behavior of one bank exerts on others.[18]

**4.2. Determinants of Systemic Risk: the Effect of Banks' Holdings of Derivatives**

---

[16] To quantify these similarities, we estimate pairwise correlations between the individual VaR and the systemic risk measure for each bank. The average correlations are 0.98, 0.94 and 0.95 for the ΔCoVaR, ΔCoES and asymmetric CoVaR, respectively.

[17] We estimate the average correlation between the GSV and the VaR for each of the 95 banks. The average correlation for the period 2002-2011 is equal to 0.98 while this correlation drops to 0.75 using the NSV.

[18] For this reason, we repeat the analysis in Subsection 5.1 using GSV as an alternative systemic risk measure.

Figure 1 depicts the average fair value (on the left-hand-side) and notional amount (right-hand-side) of the banks holdings of interest rate, foreign exchange, credit, equity and commodity derivatives over total assets. These ratios are expressed in percentages and are lagged one period (*t-1*) while the average systemic risk measure defined as the Net Shapley Value, which is contained in each panel in the figure, is depicted at period *t* such as they appear in Equation (1). Note that the y-axes have different scales for the different derivatives for a better view of their trend. According these scales we observe that interest rate derivatives represent the most widely used derivatives during the whole sample period, followed by the foreign exchange and credit derivatives. In comparison, the size of equity and commodity derivatives is almost negligible.

Overall, we observe that banks increase their notional position in derivatives during the economic boom of 2004-2006 while the fair value of holdings decreases or remains constant. This discrepancy is especially relevant in the interest, foreign exchange and credit derivatives. After the Lehman Brothers' collapse, the notional position in derivatives dramatically fell down while the fair values pick up. This increase was especially noticeable in the case of the interest rate, foreign exchange, and credit derivatives.

Focusing on the relationship between the fair value of derivatives holdings and the systemic risk, we observe that the interest rate and commodity derivatives holdings depict a downward trend one quarter before the date corresponding to the most pronounced increase in systemic risk. Nevertheless, equity holdings remained stable during this systemic episode. The correlation between the holdings of interest rate and equity derivatives lagged one quarter on the one hand and the systemic risk measure from the end of 2007 to the beginning of 2009 on the other hand; are negative and it is almost zero for case of the commodity derivatives. Finally, we find a closer relation between systemic risk and the positions in both credit and foreign exchange derivatives. We observe a slight increase in the holdings of the former and a significant increase in the latter, one quarter before the main systemic event in the sample. Thus, the correlations of the holdings of these derivatives lagged by a quarter and the systemic risk measure during the period in which we observe the highest banks contributions to systemic risk were significantly positive.

The three research questions stated in Section 2 are addressed by means of Table V, which shows the results of the estimation of Equation (1) (the baseline specification). Column 1 reports the estimated coefficients and their standard errors. Column 2 contains the economic impact of the statistically significant variables. The economic impact is defined as the standardized coefficient (i.e., the regression coefficient as in Column 1 times standard deviation of the corresponding explanatory variable) over the mean of the dependent variable. Column 3 reports the estimated coefficients when the holdings of derivatives are not used in the estimation. Column 4 contains an alternative estimation of the baseline model in which the coefficients are estimated on the basis of a fixed-effect regression at bank level with Driscoll and Kraay (1998) standard errors which are robust to heteroskedasticity, autocorrelation and correlated panels. We consider fixed effect because it enables us to control by bank specific factors that are not considered in the baseline specification. The use of Driscoll and Kraay (1998) standard errors is motivated by the properties of the residuals.

There is a significant relation between the credit, interest rate, foreign exchange and commodity derivatives holdings of bank $i$ in quarter $t$ and the contribution to systemic risk of bank $i$ in period $t+1$. Equity derivatives holdings do not affect systemic risk. Holdings of credit and foreign exchange derivatives have an increasing effect on systemic risk whereas holdings of interest rate and commodities derivatives have a decreasing effect. Foreign exchange derivatives have the highest economic impact on systemic risk.

The positive and significant effect of credit derivatives may be due to the fact that banks positions in credit derivatives are held for trading activities, either selling or buying protection, rather than for hedging loans (Minton, Stulz, and Williamson, 2009). These authors estimate that the net notional amount of these derivatives that is used for hedging loans is below 2% of the total notional amount of this type of derivatives and is less than 2% of their loans. In this line, Kiff, Elliot, Kazarian, Scarlata, and Spackman (2009) state that a large portion of CDS buyers do not hold the underlying bond but are either speculating on the default of the underlying reference or protecting other interests. The concern of heightened counterparty risk around the Lehman Brothers collapse could also help to explain this effect.[19] As pointed out by Giglio (2011), the buyer of protection could suffer even larger loses if the default of the reference entity triggers the default of the counterparty (double default), given that the buyer would have a large amount owed by the bankrupt counterparty. Even the presence of collateral may not be enough to solve this counterparty risk related to double default problem. In this context, the buyers of CDS were aware of this residual counterparty risk and considered that the best way to reduce it was to buy additional CDS protection against their counterparty.

The positive and significant effect of the variable referring to the use of foreign exchange derivatives casts some doubts on the argument against increased regulation of the foreign exchange derivatives based on the assumption of the high level of transparency of the foreign exchange market and that they performed smoothly during the financial crisis. An extreme situation, such as the devaluation of the currency of a large country, could lead to high losses for important players in this market and could make the global shock that this devaluation would cause even worse. According to the BIS (2008) report on the progress in reducing foreign exchange settlement risk, the establishment and growth of the CLS Bank has achieved significant success however, a notable share of foreign exchange transactions are settled in ways that still generate significant potential risks across the global financial system and so, further action is required. However, the clearing process is concentrated in one clearing house (the CLS Bank) and this fact could have negative systemic implications (see Duffie and Zhu, 2011). Other argument explaining this positive and significant effect can be found in Fan, Mamun, and Tannous (2009). These authors suggest that the reduction in risk gained from using foreign exchange derivatives for hedging purposes is offset by the increase in trading activities. Thus, banks could use this type of derivatives to hedge foreign exchange risk and be engaged in trading activities which would expose them to additional risk at the same time.

In regards to the negative and significant effect of the holdings of interest rate derivatives; Stulz (2004) states that derivatives can create risk at a firm level if they are used episodically and with no experience in their use. However, interest rate derivatives are broadly used by banks. The most common interest rate derivative is based on swaps, which account for around 70%, and in particular the "plain vanilla" interest rate swap. Additionally, previous literature such as Hirtle (1997), Brewer, Minton, and Moser (2000) and Carter and Sinkey (1998) suggest the use of these derivatives being more frequent in banks more exposed to interest rate risk. This result could be reflecting that derivatives enhance interest rate risk exposure for bank holding companies. In fact, banks mainly lend to firms using floating rates and for this reason, they could aim to increase their positions in interest rate derivatives when the interest rates begin to diminish. Thus, we find that the correlation between the 3-month LIBOR (10-year U.S. Government bond yield) and the holdings of interest rate derivatives is -0.75 (-0.91) indicating that the use of these derivatives is determined by decreases in the interest rate. This finding is in line with the one presented by Christoffersen, Nain, and Oberoi (2009) who show a negative relation between the use of interest rate derivatives and the interest rate movements. These authors argue that even if companies are able to anticipate the interest rate policy, it is possible that they cannot adjust the debt exposure; however they can adjust the swap exposures to reduce the cost of debt. Carter and Sinkley (1998) and Downing's (2012) results also support the hypothesis that banks use interest-rate derivatives to hedge interest rate risk. The negative correlation could also be consistent with a higher cost of interest rate volatility during economic downturns.

---

[19] Counterparty risk is of special relevance given that the credit risk of a given company can spill over to other companies leading to industry and market-wide increases in the CDS spreads (Jorion and Zhang, 2007, 2009).

The effects of the use of equity and commodity derivatives on banks' risk or performance have been scarcely addressed in previous literature. One reason explaining the lack of empirical studies on this topic could be the lower relative importance of the positions on equity and commodity derivatives as can be observed in Figure 1. However, while the effect of the equity derivatives is not significant at any standard significance level, commodity derivatives have a negative effect on the dependent variable with a p-value slightly higher than 0.10. The holdings of commodity derivatives, as occurs with the other derivatives, could be justified by the search for higher yields in a low interest rate environment. Moreover, the increase in the use of commodity derivatives could be propitiated, as stated in Basu and Gavin (2010), by the movement from real estate derivatives to commodity derivatives coinciding with the appearance of the problems in the subprime market. Other theories suggest that banks could use commodity derivatives to hedge inflation risk, to take advantage of the increase in the commodity prices around the systemic event, or because they are negatively correlated with equity and bond returns (Gorton and Rouwenhorst, 2006). Basu and Gavin (2010) show that when commodity prices peak in June 2008, the correlation with the equity index was, on average, negative. In fact, we observe the highest holdings of commodity derivatives by banks in this period. After summer 2008 the correlation becomes extremely positive and holdings of commodity derivatives diminished substantially from their highest levels.

Column 3 reports the results obtained when we estimate a restricted model in which we do not include the holding of derivatives as explanatory variables. On the basis of these results and the ones obtained in Column 1 we run an *F-test* of joint significance of the derivatives coefficients. The test rejects the null hypothesis that states that the coefficients for all the derivatives related variables are zero. This result indicates that the derivatives holdings are jointly significant and should not be omitted from the Equation (1). The results under the alternative fixed-effects methodology reported in Column 4 are fully in agreement with the ones obtained under the Prais-Winsten regression apart from the total loans that are not significant in the new specification. So, our results seem robust to the introduction of bank fixed effects when considering the properties of the residuals.

Regarding the effect of the size, substitutability, interconnectedness and balance sheet related variables, we find that increases in the following variables increase systemic risk contributions: total loans, net balance to banks belonging to the same banking group, leverage ratio and the proportion of non-performing loans over total loans. On the other hand, increases in total deposits decreases systemic risk. The effect of the size related variables is not significant after controlling by this battery of variables.[20] The variables with the highest economic impact on systemic risk are the proportion of non-performing loans to total loans and the leverage ratio. For instance, one standard deviation increase in the proportion of non-performing loans to total loans in quarter *t*, increases the bank's contribution to systemic risk in quarter *t+1* to 17% above its average level.

No other variable presents significant effects. In particular and in contrast to Brunnermeier et al. (2011) non-interest to interest income is not significant when derivatives holding are included in the equation. This discrepancy could be also due to the different sample, time periods, systemic risk measures, or explanatory variables employed in the two papers. Finally, the aggregate systemic risk level in the previous quarter contributes positively and significantly to increase the individual contribution to systemic risk but the effect of aggregate systemic risk does not go beyond one quarter before the current one.[21]

---

[20] We have repeated the analysis using the logarithm of total assets and its square as alternative variables to proxy the bank size and find similar results.

Summing up, although the two variables with the highest economic impact on the bank's contribution to systemic risk are the non-performing loans relative to total loans and the leverage variables; the bank's holdings of financial derivatives also have significant effects but of a much lower magnitude.

Some literature has considered that the use of derivatives should not pose significant levels of risk to the economy or to individual corporations. For instance, Stulz (2004) concludes that we should not fear derivatives but have a healthy respect for them. He considers that losses from derivatives are localized but the whole economy gains from the existence of derivatives markets. Hentschel and Kothari (2001) question whether corporations are reducing or taking risks with derivatives, their answer is "typically not very much of either". The authors find an absence of higher risks due to the effect of derivatives (even among firms with large derivatives positions) which in their view shows that the concern over widespread derivative speculation is unfounded. Along this line, Cyree, Huang, and Lindley (2012) find that the effects of derivatives (interest rate, foreign exchange, and credit derivatives) on market valuation are not statistically distinguishable from zero in either good times or bad times.

Our results do not imply that the use of derivatives by banks is inconsequential as far as systemic risk is concerned. They do imply that their impact, albeit statistically significant, plays a second fiddle in comparison with traditional variables such as leverage or the proportion of non-performing loans over total loans. Furthermore, the use of derivatives could indirectly affect the systemic contribution of banks given that derivatives require limited up-front payments and enable banks to take more leveraged positions. Additionally, the use of derivatives could lead to diminished monitoring of loans when the banks are considered to have used the right hedging strategies. In this respect, Morrison (2005) finds that credit derivatives can reduce banks' incentives to monitor their loan portfolios.

---

[21] The use of these lagged measures enables us to mitigate the potential autocorrelation in the residuals. Nevertheless, we check whether there is significant first order autocorrelation in the residuals by means of individual tests for each bank. The coefficient for the first order autocorrelation is only significant in 25 out of the 95 banks being its average magnitude around 0.3 for these 25 banks. We conduct an additional test to discard the existence of first order correlation in the residuals. Thus, we calculate the average residual for each date across the 95 banks and regress this series on its lagged value. The estimated coefficient is not significantly different from zero and so, we do not find evidence in favor of the presence of autocorrelation.

**4.3. Dealing with Endogeneity**

The baseline model directly relates the banks' derivatives activity to their individual contribution to systemic risk and hence, it may potentially suffer from endogeneity problems in the sense that banks with certain levels of systemic risk or banks with other specific characteristics may self-select them into derivatives participation. To address this important issue we perform four different but complementary endogeneity analyses.

The first analysis relies on a Heckman two-stage model as in Shao and Yeager (2007) and Fung et al. (2012). In the first stage we estimate a Probit regression model to examine how the likelihood of using a certain type of derivative is affected by bank-specific characteristics and the total holdings of the remaining four types of derivatives. From this first-stage selection model we get the inverse Mills ratio that is a self-selection parameter that will be used as an input in the second-stage regression. In the second-stage we estimate the baseline equation (Equation 1) including as an additional variable the self-selection parameter and excluding the four types of derivatives that were used in the first-stage

To conduct the first-stage we consider several variables employed in the baseline analysis as well as other additional variables related to the bank trading activity. Thus, we use the two main determinants of the bank contribution to systemic risk as suggested by the baseline analysis: leverage and non-performing loans. These two variables are closely related to the bank's on-balance sheet credit risk and the costs of financial distress. As pointed out by Minton et al. (2009) banks that are more likely to bear greater costs of financial distress are expected to hedge more. Additionally, through hedging, banks can also take on more risks. We expect that the higher the leverage or the amount of non-performing loans, the higher should be the use of derivatives.

Bank size is measured by means of the logarithm of bank market value. According to Sinkey and Carter (2000), if economies of scale or scope exist in bank' derivatives activities; a positive relation should exist between the use of the different types of derivatives and the size. Whidbee and Wohar (1999) also suggest that small firms may lack the technology and expertise to effectively use derivatives to manage their risk exposure.

The loans related variables capture the exposure to unexpected changes in interest rates and to loan counterparty default. For this reason, we use the total loans relative to total assets as a potential driver for the use of interest rate and credit derivatives. In the case of foreign exchange derivatives we use a dummy indicating if banks have granted loans to foreign institutions (banks and Governments) to consider the direct effect of loans in foreign currency. In the same vein, we consider the dimension of the loan activity that could affect the use of commodity derivatives and employ the loans for financing agricultural production purposes. The effect of loans on the use of derivatives is not obvious. In fact, Minton et al (2009) show that banks do not use credit derivatives to hedge loans.

The correlation with the S&P500 and the bank's commercial paper are used to measure the interconnection outside the group through the stock market and the short-term funding market, respectively. The interconnectedness through the commercial paper market is especially relevant because the higher the dependence on short-term funding, the higher is the exposition to financing stress. This exposition could explain the use of derivatives either for hedging exposures or to get additional profits. The effect of this variable is also related to liquidity availability. A low dependence on the commercial paper would be an indicator of a high degree of liquidity. According to Carter and Sinkey (1998), liquidity could be employed as an alternative source for hedging that makes the participation in derivatives market less frequent. For all the previous reasons, we expect a positive effect of the short-term funding needs and also the exposure to the stock exchange on the use of derivatives holdings.[22]

We use the bank's non-interest to interest income as a measure of the bank non-core relative to the bank traditional lending activities. Banks with a high ratio would participate more in trading activities and so, in derivatives markets. According to Carter and Sinkey (1998), another measure of the potential costs associated with participation in the derivatives market is whether a bank uses other derivative instruments. Banks that invest in the human capital and internal control systems necessary to be active in the market for derivatives are more likely to use more than one type of derivative. Thus, we use as an additional explanatory variable the total holdings of the other four types of derivatives relative to total assets.[23] Finally, as in Cyree, Huang, and Lindley (2012) we use time-effects to control for the effect of other global factors that could affect to the banks' decision of using derivatives.

---

[22] The commercial paper relative to total assets is excluded from the regression analyzing the use of interest rate derivatives because it perfectly predicts success.

The effect of the determinants of the use of derivatives (first-stage) and the determinants of systemic risk including the self-selection parameter (second-stage) are reported in Table VI. As in Sinkey and Carter (1998) we find that consistent with the avoidance of the costs of financial distress, banks with high leverage ratios and large proportions of non-performing loans are more likely to use all kind of derivatives. In line with Minton et al. (2009), who do not find a difference in the amount of non-performing loans of the net buyers and the non-net buyers of protection, we obtain that non-performing loans do not affect significantly to the use of credit derivatives. Not only the non-performing loans but total loans are also a significant driver for the use of all derivatives with the exception of equity derivatives. In line with previous research on banks' use of derivatives, we find that an increase in bank size is positively and significantly associated with the likelihood of using any type of derivative. In view of the positive and significant coefficient for the holdings of the other types of derivatives we confirm Carter and Sinkey (1998), Fung et al. (2012) and Minton et al.'s (2009) result: the degree of expertise in other derivatives lead to an increase in the holdings of a given derivative. The positive relation between non-traditional or trading activities and the use of derivatives is supported by the significant coefficients of the non-interest to interest income for the use of credit, interest rate, and equity derivatives. The variables related to the bank's interconnectedness are significant for some derivatives with the expected sign, but this is not the case for commodity derivatives. In sum, banks using derivatives can be characterized as highly leveraged large banks, with high ratios of loans to total assets and non-performing loans, very active in other derivatives markets, and with short-term funding needs. The significant effects of the variables related to the bank's risk and interconnectedness could also help us to explain the trends in the holdings of derivatives during the crisis as reported in Figure 1.

---

[23] As an alternative to the total holdings of other derivatives to measure the banks experience in derivatives markets we analyze the effect of banks acting as primary dealers on their use of derivatives as in Sinkey and Carter (2000). So, we use a dealer dummy to control for their actions in the derivatives market such that it is coded as 1 for the ISDA primary members and 0 otherwise. We find similar results for the two proxies of banks experience such that primary dealers also hold a larger proportion of derivatives.

Regarding the results obtained in the second stage, we confirm that endogeneity is not a problem in our analysis. The holdings of foreign and credit derivatives significantly increase the bank's contribution to systemic risk while the holdings of interest rate derivatives exhibit a negative level but non-significant at 5% level. Nevertheless, the inverse Mills ratio is negative and significant at 5% suggesting that banks that are considered to have ex ante a lower contribution to systemic risk are more likely to use interest rate derivatives. The other two types of derivatives do not have a significant effect at any standard level of significance.

The previous results rely on the definition of the model explaining the use of derivatives by banks. So, to make results independent of the model specification we perform a second test in which we do not impose any determinant of the decision for holding derivatives but simply use as explanatory variable the derivative holdings lagged one period. In this analysis, which is based on Westerlund and Narayan (2012), we first compute the adjusted systemic risk measure (AdjSR) that is the residual obtained after regressing systemic risk on all the explanatory variables employed in the baseline analysis with the exception the derivatives variables. We next perform the following panel data predictive regression model individually for each derivative type *j*:

$$\text{AdjSR}_{i,t} = \alpha_j + \beta_j \text{DH}^j_{i,t-1} + \varepsilon^j_{i,t} \qquad (3)$$

where $\text{DH}^j_{i,t-1}$ denotes the derivatives holdings of type *j* which is a predictor variable with the following autoregressive representation:

$$\text{DH}^j_{i,t} = \delta_j + \rho_j \text{DH}^j_{i,t-1} + v^j_{i,t} \qquad (4)$$

The correlation between the residuals of the two models ε and ν captures, essentially, the extent of endogeneity of the predictor variable $\text{DH}^j_{i,t-1}$ and can be estimated running the following regression:

$$\varepsilon^j_{i,t} = \tau_j + \omega_j v^j_{i,t} + \varphi^j_{i,t} \qquad (5)$$

Testing whether $\omega_j$ is different from zero would enable us to assess the extent to which endogeneity is a problem in our study. If it turns out that endogeneity is in fact a problem, then we follow Westerlund and Narayan (2012) and run the following panel regression robust to heteroskedasticity to obtain consistent estimations for the effect of derivative holdings:

$$\text{AdjSR}_{i,t} = \theta_j + \beta_j \text{DH}_{i,t-1}^j + \psi_j\left(\text{DH}_{i,t}^j - \hat{\delta}_j - \hat{\rho}_j \text{DH}_{i,t-1}^j\right) + \eta_{i,t}^j \qquad (6)$$

where $\hat{\rho}_j$ and $\hat{\delta}_j$ are the estimated coefficients in Equation (4). Notice that we estimate Equation (6) using simultaneously all the derivatives holdings with problems of endogeneity while the derivatives that are found to be exogenous are used to filter the adjusted systemic risk measures. The results are reported in Column 1 of Table VII. For the sake of brevity we only report the coefficients for the derivatives in which we found potential problems of endogeneity: credit, foreign exchange, and interest rate derivatives. The results confirm the ones obtained in the baseline specification: positive (negative) and significant coefficients for the holdings of credit and foreign exchange derivatives (interest rate derivatives).

As a third way of dealing with the potential endogeneity problem, we use an instrumental variable approach. We focus on the three types of derivatives with potential problems of endogeneity according to the results obtained in the previous analysis: credit, foreign exchange, and interest rate derivatives and use two groups of instruments. As a first group of candidates for instruments we consider bank information that is related to its trading activity. According to Norden et al. (2012), banks typically start hedging activities in derivatives following trading in derivatives. Moreover, Minton et al. (2009) find that the use of credit derivatives is highly correlated with the trade of other derivatives. In this study we use the assets in trading accounts lagged one quarter and the total fair value of derivatives held for trading, as in Norden et al. (2012), also lagged one quarter.[24] The channel supporting the use of these instruments is that the larger the bank's trading activity, the higher are the bank's holdings of derivatives. As a second group of candidates for instruments we consider the expertise of the bank in the different types of derivatives that were found to be potentially endogenous in the previous analysis. We define the degree of expertise in each derivative type as the ratio of face value of the holdings on such derivative relative to the sum of the face values of the five types of derivatives used in this study. This ratio is lagged four periods (one year). The higher the ratio, the more likely the bank will hold such derivative type in the future. Given that this variable is defined relative to the total holdings of derivatives it is not related to size and it is not related to the total activity in derivatives but mainly to the expertise of the bank in a given type of derivatives relative to the other types of derivatives.

---

[24] The Bank Holding Company Data contains the category derivatives held for trading for all derivatives except credit derivatives.

We run the instrumental variable regression with fixed effects and robust to heteroskedasticity in which the holdings of the three types of derivatives are instrumented through the two previous groups of instruments. The remaining explanatory variables are the ones employed in the baseline analysis including the fair value holdings of the other two derivatives that were not found to be endogenous. Column 2 of Table VII reports the coefficients for the three potentially endogenous derivatives types. We perform the Kleibergen-Paap Rank LM statistic to check whether the equation is identified that is, whether the excluded instruments (the two group of variables explained above) are "relevant" (correlated with the endogenous regressor). According to this underidentification test we reject the null hypothesis (equation is underidentified) and so, the instruments are relevant and the model is identified. We also perform a weak identification test to analyze whether the trading activity and expertise related variables are correlated with the holdings of derivatives but only weakly. For such aim we use the Kleibergen-Paap Wald Rank F statistic according to which we reject the hypotheses that the equation is weakly identified. Finally, we test the validity of the instruments on the basis of the overindentification test based on Hansen J statistic and confirm that the instruments are valid given that they are uncorrelated with the error term. As it can be inferred from the significant coefficient for the three types of derivatives holdings we conclude that the potential endogeneity of these holdings does not bias our results. Although non-reported, the other two types of derivatives do not exhibit a significant effect while the signs and degree of significance of the remaining variables are similar to the ones obtained in the baseline analysis.

As the final attempt to analyze if the baseline results may be distorted by endogeneity and following Brunnermeier et al. (2011), we consider the bankruptcy filing of Lehman Brothers on September 15, 2011 (2008Q3) as an exogenous shock. We employ a difference-in-differences approach to analyze whether banks with different levels of derivatives holdings contribute differently to systemic risk measures when they face the unexpected shock of Lehman Brothers bankruptcy. Thus, banks with more holdings of derivatives are defined as the treatment group, and banks with fewer holdings are the control group. As in Brunnermeier et al. (2011) we rank all commercial banks based on their holdings of each type of derivative in the year 2007Q2 and Q3 (average over the two quarters of the ratio of derivatives holdings relative to total assets). The dummy variable of *Top-quartile* is set to unity if the bank's holdings of a given type of derivative is in the top-quartile (75-percentile and above), and zero if it is in the bottom-quartile (25-percentile and below).[25] The dummy variable of *Post-Lehman* is set to unity if the date is 2008Q4 (the quarter after the bankruptcy filing of Lehman Brothers), and zero if the date is 2007Q4 (one year before the bankruptcy filing of Lehman Brothers). A third dummy variable *Top-quartile\*Post-Lehman* is the cross-product of the previous two dummy variables. This analysis is immune to the abrupt change in market conditions since the third quarter of 2007 and the results obtained in this analysis do not depend on the trends in derivatives usage observed during the crisis.

The results of the difference-in-difference regression for those holdings of derivatives with potential problems of endogeneity are reported in Table VIII. The control variables employed in the baseline analysis are also used in this analysis. The coefficients of the interaction dummy (*Top-quartile\*Post-Lehman*) are significantly positive for the credit and foreign exchange derivatives. It confirms that banks that before the Lehman episode had a large amount of the two previous types of derivatives relative to total assets contributed more to systemic risk after the shock of Lehman than those banks with lower relative holdings of those derivatives. The holdings of interest rates derivatives do not exhibit a significant effect on systemic risk after the shock of Lehman's collapse. On the contrary, the holdings of interest rate derivatives have helped to reduce the banks' contributions to systemic risk before and also during the crisis given that the dummy *Top-quartile* is significantly negative. As expected in the three cases, the Lehman bankruptcy leads, on average, to a larger contribution to systemic of the banks in the sample.

## 5. Robustness Test

### 5.1. Alternative Indicators of Systemic Risk

---

[25] Note that in the control group we have more than 25% of the banks when the proportion of banks that do not hold a given type of derivative is larger than 25%.

We first study whether the sign and significance level of the holdings of derivatives differs when we employ alternative measures of the individual bank's contribution to the systemic risk. To do that we estimate the baseline specification considering the following measures: (1) alternative specification of the NSV in which we include a synthetic bank constructed as the weighted average of the remaining banks that do not belong to the system and are not used to estimate the measure; (2) variation of the NSV in which we aggregate the information within a given quarter by summing up all the weekly estimated measures instead of using the end of quarter information; (3) GSV; (4) $\Delta CoVaR$; (5) $\Delta CoES$; (6) Asymmetric $\Delta CoVaR$). Results are summarized on Table IX. In the interest of brevity, we only report the effects of derivatives. Note that due to the different nature of each measure of systemic risk the magnitude of the coefficients is not comparable across the different columns.

We observe a consistent significant effect of credit, interest rate, and foreign exchange derivatives across the alternative systemic risk measures. Independently on the measure, the use of credit and foreign exchange derivatives leads to higher contributions to systemic risk while the use of interest rate derivatives leads to lower contributions. Commodity derivatives are only significant when we use the GSV measure while equity derivatives never exhibit an effect significantly different from zero. This confirms the relevance of derivatives holdings in explaining the individual contribution of banks to systemic risk.

## 5.2. Alternative Model Specifications

### 5.2.1. Increasing the number of lags of explanatory variables

Banks make their operating and financial decisions at time *t* and these decisions generate profits, or returns, between *t* and *t+1*. To avoid any potential simultaneity bias we lag all the variables (including the holdings of derivatives) 2, 3, and 4 quarters before the date in which systemic risk is measured. The results show a similar effect, in terms of the sign and the significance of the explanatory variables, to the one observed in the baseline specification. Interestingly, under these new specifications we document the existence of a significant U-shaped relationship between size and systemic risk such that Bank of America, Citigroup, JP Morgan Chase and Co and Wells Fargo and Co. are located at the right-hand-side of the vertex (i.e., in the upward trend).

**5.2.2. Different specifications for derivatives holdings**

We next check the impact of derivatives holdings in the Equation (1) using four alternative specifications to handle those holdings.

The period before Lehman's collapse exhibited high liquidity, low interest rates, and possibly some investor's overconfidence; all these factors might lead to some misvaluation of derivatives prices and their risks. To avoid any potential bias derived from the use of the derivatives' fair value; we consider instead the notional amount to value the holdings of derivatives and find similar results confirming the robustness of our baseline results.

Secondly, we discretize the derivative holdings variables and consider dummy variables for each type of derivatives that take value 1 when the fair value of the bank's holdings is above the third quartile in the distribution of the banks' holdings and takes value zero otherwise. The resultants new variables do not reflect the banks' holdings used to minimize the risk or to maximize the profit; they simply reflect the use or otherwise of derivatives. We again find a positive and significant effect of foreign exchange and credit derivatives.

The use of derivatives could be influenced by the changes in the economic situation and be used as a response to these changes such that there could be a problem of reserve causality. To discard that this affects our results we design a two-stage test. In the first stage, we filter the holdings of each type of derivative using the same set of global factors (i.e., S&P500 return, commodity index return, U.S. price home index return, VIX, U.S. LIBOR, Euro/Dollar exchange). In the second stage, we use the filtered residuals for each derivative as the explanatory variables proxying the "idiosyncratic" holdings of derivatives and estimate the baseline specification. The results are similar to the ones obtained in the baseline specification: holding of credit and foreign exchange derivatives contribute to a significant increase of systemic risk whereas holdings of interest rate derivatives exhibit the opposite effect.

Finally, to provide some intuition about the impact of derivatives activity on the overall systemic risk, we aggregate all the variables by date and run the baseline regression for the 45 quarters in the sample. To ensure some degrees of freedom we only consider the significant variables in Table V apart from the derivatives holdings. We obtain that the aggregate levels of credit derivatives (interest rate and commodity derivatives) of the commercial banking industry significantly increase (decrease) the overall level of systemic risk while foreign exchange and equity derivatives do not exhibit a significant effect.

### 5.2.3. Alternative control variables and specifications

As in Brunnermeier et al. (2011) we also use as an explanatory variable the lagged level of bank risk according to its VaR (defined in positive terms) instead of the aggregate lagged level of systemic risk. In this case, the R-squared increases from 0.49 to 0.53 and the effect of the VaR variable is positive and significant at any level of significance. The effect of the remaining explanatory variables is similar to those in the baseline regression. In view of this, our results are robust to the use of the bank's VaR to control for the level of risk in the previous quarter.

To take into account the effect of the degree of concentration in the banking sector, we include the Herfindahl-Hirschman index referred to the banks' total assets as an additional explanatory variable. This variable does not have a significant effect at any level of significance and both the coefficients and levels of significance of the explanatory variables are unchanged with respect to the results obtained in the baseline regression. [26]

The explanatory variables are defined as ratios over total assets to correct the firm size effect and possible non-linearity. Alternatively, we use two different strategies to handle the non-linearity in the explanatory variables. The first strategy relies on linearizing the ratios by taking the logarithm of the ratio plus one. The second strategy relies on applying the Fisher transformation to the independent variables. This transformation makes the distribution for the transformed variables closer to normality and improves its degree of symmetry in case they were skewed. However, we can only consider this transformation for those variables ranging between –1 and 1 and hence, we cannot linearize neither the leverage nor the non-interest to interest income variables. The results of those strategies are similar to the ones reported in the baseline analysis.

## 6. Conclusions

---

[26] Detailed results for all the extensions and robustness tests carried out in this section are available upon request.

The recent financial crisis has exposed the dangers lurking in oversized banking sector balance sheets. One major concern for regulators has been the astonishing growth in derivatives markets and consequently in the swelling of derivatives holdings in banks' balance sheets. The aim of this paper is to address the extent to which this situation has increased systemic risk.

First, we propose an alternative measure of the individual contribution to systemic risk that is based on the Gross Shapley Vale and that we call Net Shapley Value. This measure allows us to get rid of the idiosyncratic component present in the last measure. Then, we compare alternative systemic risk measures and find that the Net Shapley Value outperforms the others. Using the Net Shapley Value as our proxy for systemic risk we find strong evidence of derivative holdings acting as leading indicators of banks' systemic risk contributions. However, their effects are not alike because credit and foreign exchange derivatives have a positive effect on systemic risk whereas holdings of interest rate derivatives have a negative effect. Given that the baseline model directly relates the banks' derivatives activity to their individual contribution to systemic risk, the results may potentially suffer for endogeneity problems in the sense that banks with certain specific characteristics self-select them into derivatives participation. To address this concern we provide four pieces of evidence that enable us to confirm that our findings come indeed from a causal relation between banks' derivatives holdings and their contribution to systemic risk.

Finally, other balance sheet variables are also leading indicators of systemic risk contributions. Increases in the following variables increase systemic risk contributions: total loans, net balance to banks belonging to the same banking group, leverage ratio and the proportion of non-performing loans (measured in this case relative to total loans), on the other hand, increases in total deposits decreases systemic risk. The variables with the highest economic impact on systemic risk are the proportion of non-performing loans to total loans and the leverage ratio. In fact, in terms of economic impact on systemic risk, the balance sheet items related to traditional banking activities (leverage, non-performing loans) have the stronger effect.

Our results provide some implications for regulators and bankers alike. Since the burst of the subprime crisis derivatives have been considered as a relevant factor in the destabilization of the banking system. Nevertheless, the claims that all derivatives have pernicious effects on the overall financial system are not borne out by the data. Therefore, the process of re-regulation that is under way in many countries should be carefully designed to avoid hindering activities that are actually diminishing systemic risk. Financial stability is a public good that can inform corporate investment and financing decisions and thus any new regulatory initiative should be very carefully designed to give the different instruments within an asset class, in this case, derivatives, the appropriate regulatory oversight. This "recommendation" mainly concerns developed markets and well regulated banking systems, while emerging markets may face different challenges.

On the other hand, given the empirical evidence reported in this paper, the economic impact of non-performing loans and leverage on systemic risk is much stronger than the derivatives' impact. Therefore the traditional banking activities related to these two items should be closely watched by regulators worried about systemic risk episodes.

**References**

Acharya, V. V., 2011, "A Transparency Standard for Derivatives," in M. Brunnermeier and A. Krishnamurthy, (eds.) "Risk Topography: Measuring Systemic Risk," NBER.

Acharya, V.V., L.H. Pedersen, T. Philippon, and M. Richardson, 2011a, "Quantifying Systemic Risk: How to Calculate Systemic Risk Surcharges," Chapter in NBER Book Quantifying Systemic Risk, Joseph G. Haubrich and Andrew W. Lo, editors.

Acharya, V.V., L.H. Pedersen, T. Philippon, and M. Richardson, 2011b, "Measuring Systemic Risk," Working paper, New York University.

Acharya, V.V., and A. Thakor, 2011, "The Dark Side of Liquidity Creation: Leverage and Systemic Risk," Working Paper, Federal Reserve Bank of New York.

Adrian, T., and M.K. Brunnermeier, 2011, "CoVar," Federal Reserve Bank of New York Staff Report 348.

Allen, L., T.G. Bali, and Y. Tang, 2012, "Does Systemic Risk in the Financial Sector Predict Future Economic Downturns?" *The Review of Financial Studies* 25, 3000-2036.

Bank for International Settlements, 2008, "Progress in Reducing Foreign Exchange Settlement Risk."

Basu, P., and W.T. Gavin, 2010, "What explains the growth in commodity derivatives?" Federal Bank of St. Louis review 93, 37-48.

Behn, B., J.H. Choi, and T. Kang, 2008, "Audit Quality and Properties of Analysts' Earnings Forecasts," *The Accounting Review* 83, 327-349.

Boyson, N., J. Helwege, and J. Jindra, 2011, "Crises, Liquidity Shocks, and Fire Sales at Financial Institutions," In Midwest Finance Association 2012 Annual Meetings Paper.

Brewer, E., B. Minton, and J. Moser, 2000, "Interest-Rate Derivatives and Bank Lending," *Journal of Banking and Finance* 24, 353-379.

Brunnermeier, M.K., G. Dong, and D. Palia, 2011, "Banks' Non-Interest Income and Systemic Risk," Princeton University and Rutgers University.

Calmès, C., and R. Théoret, 2010, "The impact of off-balance-sheet activities on banks returns: an application of the ARCH-M to Canadian data," *Journal of Banking and Finance* 34, 1719–1728.

Carter, D.A., and J.F. Sinkey, 1998, "The Use of Interest Rate Derivatives by End-Users: The Case of Large Community Banks," *Journal of Financial Services Research* 14, 17-34.

Christoffersen, P., A. Nain, and J. Oberoi, 2009, "Interest Rate Shocks and Corporate Risk Management," Working Paper, McGill University

Cummins, J.D., and M.A. Weiss, 2010, "Systemic Risk and the U.S. Insurance Sector," Temple University Working Paper.

Cyree, K., P. Huang, and J. Lindley, 2012, "The Economic Consequences of Banks' Derivatives Use in Good Times and Bad Times," *Journal of Financial Services Research* 41, 121-144.

Dell'Ariccia, G., L. Laeven, and D. Igan, 2008, "Credit Booms and Lending Standards: Evidence from the Subprime Mortgage Market," IMF Working Paper 08/106, Washington, DC.

Dodd-Frank Wall Street Reform and Consumer Protection Act.

Downing, J., 2012, "Banks, Price Risk, and Derivatives: Evidence and Implications for the Volcker Rule and Fair-Value Accounting," Working Paper.

Driscoll, J. C. and A. C. Kraay, 1998, "Consistent Covariance Matrix Estimation with Spatially Dependent Panel Data", *Review of Economics and Statistics* 80, 549-560.

Duffie, D., and H. Zhu, 2011, "Does a central clearing counterparty reduce counterparty risk?" *The Review of Asset Pricing Studies* 1, 74-95.

Fan, H., A. Mamun, and G. Tannous, 2009, "What determines Bank Holding Companies Foreign Exchange Derivatives For Trading and for Other-Than-Trading," Working Paper.

Ferreira, M.A., and P.A. Laux, 2007, "Corporate Governance, Idiosyncratic Risk, and Information Flow," *Journal of Finance* 62, 951-989.

Fung, H.G., M.M. Wen, and G. Zhang, 2012, "How Does the Use of Credit Default Swaps Affect Firm Risk and Value? Evidence from US Life and Property/Casualty Insurance Companies," *Financial Management* 41, 979-1007.

Giglio, S., 2011, "Credit Default Swap Spreads and Systemic Financial Risk," Working Paper.

Gorton, G., and K.G. Rouwenhorst, 2006, "Facts and Fantasies about Commodity Futures," *Financial Analysts Journal* 62, 47-68.

Gonzalo, J., and C. Granger, 1995, "Estimation of Common Long-Memory Components in Cointegrated Systems," *Journal of Business and Economic Statistics* 13, 27-35.

Granger, C.W.J., 1969, "Investigating Causal Relations by Econometric Models and Cross-Spectral Methods," *Econometrica* 37, 424-438.

Greenspan, A., 2005, "Risk Transfer and Financial Stability," Remarks by Chairman to the Federal Reserve Bank of Chicago's Forty-first Annual Conference on Bank Structure, Chicago, Illinois, on May 5, 2005.

Hentschel, L., and S.P. Kothari, 2001, "Are Corporations Reducing or Taking Risks with Derivatives?" *Journal of Financial and Quantitative Analysis* 36, 93-118.

Hirtle, B.J., 1997, "Derivatives, Portfolio Composition, and Bank Holding Company Interest Rate Risk Exposure," *Journal of Financial Services Research* 12, 243-266.

Hirtle, B., 2008, "Credit Derivatives and Bank Credit Supply," FRB of New York Staff.

Instefjord, N., 2005, "Risk and Hedging: Do Credit Derivatives Increase Bank Risk?" *Journal of Banking and Finance* 29, 333-345.

Jaffe, D., 2003, "The Interest Rate Risk of Fannie Mae and Freddy Mac," *Journal of Financial Services Research* 24, 5-29.

Jorion, P., and G. Zhang, 2007, "Good and Bad Credit Contagion: Evidence from Credit Default Swaps," *Journal of Financial Economics* 84, 860-883.

Jorion, P., and G. Zhang, 2009, "Credit Contagion from Counterparty Risk," *Journal of Finance* 64, 2053-2087.

Kiff, J., J. Elliot, E. Kazarian, J. Scarlata, and C. Spackman, 2009, "Credit Derivatives: Systemic Risks and Policy Options," IMF Working Paper 09/254.

Koenker, R., and G. Bassett, 1978, "Regression Quantiles," *Econometrica* 46, 33-50.

Koppenhaver, G.D., 1985, "Bank Funding Risk, Risk Aversion, and the Choice of Futures Hedging Instrument," *Journal of Finance* 40, 241-255.

Livne, G., G. Markarian, and A. Milne, 2011, "Bankers' Compensation and Fair Value Accounting," *Journal of Corporate Finance* 17, 1096-1115.

Lopez, G., A. Moreno, A. Rubia, and L. Valderrama, 2011, "Asymmetric Covar: An Application to International Banking," Systemic Risk, Basel III, Financial Stability and Regulation 2011.

Minton, B., R. Stulz, and R. Williamson, 2009, "How Much Do Banks Use Credit Derivatives to Hedge Loans?" *Journal of Financial Services Research* 35, 1-31.

Mizrach, B., 2011, "Leverage and VaR as Measures of Bank Distress: Comment on Endogenous and Systemic Risk," Rutgers University Working Paper.

Morrison, A.D., 2005, "Credit Derivatives, Disintermediation and Investment Decisions," *Journal of Business* 78, 621-648.

Nijskens, R., and W. Wagner, 2011, "Credit risk transfer activities and systemic risk: How banks became less risky individually but posed greater risks to the financial system at the same time," *Journal of Banking and Finance* 35, 1391-1398.

Norden, L., C. Silva-Buston, and W. Wagner, 2012, "Banks' Use of Credit Derivatives and the Pricing of Loans: What Is the Channel and Does It Persist Under Adverse Economic Conditions?", Working paper, Erasmus University Rotterdam and Tilburg University.

Rodriguez-Moreno, M., and J.I. Peña, 2013, "Systemic Risk Measures: the Simpler the Better?" *Journal of Banking and Finance* 37, 1817-1831.

Shao, Y., and T.J. Yeager, 2007, "The Effects of Credit Derivatives on U.S. Bank Risk and Return, Capital and Lending Structure," Working Paper.

Stulz, R.M., 2004, "Should We Fear Derivatives?" *Journal of Economic Perspectives* 18, 173-192.

Stulz, R., 2010, "Credit Default Swaps and the Credit Crisis," *Journal of Economic Perspectives*, 24, 73-92.

Sinkey, J. F., and D.A. Carter, 2000, "Evidence on the financial characteristics of banks that do and do not use derivatives", *Quarterly Review of Economics and Finance*, 40, 431-449.

Tarashev, N., C. Borio, and K. Tsatsaronis, 2010, "Attributing systemic risk to individual institutions," BIS Working Papers No 308, May.

Venkatachalam, M., 1996, "Value-Relevance of Banks' Derivatives Disclosures," *Journal of Accounting and Economics* 22, 327-355.

Westerlund, J., and P. K. Narayan, 2012, "Does the choice of estimator matter when forecasting returns?", *Journal of Banking & Finance* 36, 2632–2640.

Whidbee, D. A., and M. Wohar, 1999, "Derivative activities and managerial incentives in the banking industry", *Journal of Corporate Finance* 5, 251-276.

# Appendix A

In this appendix we provide a detailed description of the explanatory variables obtained from the database Bank Holding Company Data (Federal Reserve Bank of Chicago) that are employed in this paper:

*Fair value of credit derivatives*: this variable is defined as the sum of the total fair value (positive and negative) of the total gross notional amount in which the reporting bank is beneficiary or guarantor.

*Fair value of interest rate, foreign exchange, equity and commodity derivatives*: this variable is defined as the sum of the total fair value of the total gross notional amount for each of the four previous types of derivative contracts held for trading and for purposes other than trading by the banks. The total fair value is obtained as the sum of the positive and negative fair values.[27]

*Commercial paper*: The total amount outstanding of commercial paper issued by the reporting bank holding company to unrelated parties. Commercial paper matures in 270 days or less and is not collateralized.

*Loan to banks*: this variable includes all loans and all other instruments evidencing loans (except those secured by real estate) to depository institutions chartered and headquartered in the U.S. and the U.S. and foreign branches of banks chartered and headquartered in a foreign country.

*Maturity mismatch:* this variable is defined as the ratio of short term debt relative to total assets.

*Net balance to bank*: difference between all balances and cash due to related banks[28] and all balances and cash due from related banks. Due to accounts are liabilities accounts that represent the amount of funds currently payable to another account. Due from accounts are assets accounts that represent the amount of deposits currently held at another company.

*Net balance to non-bank:* this variable is the difference between all balances and cash due to related non-banks and all balances and cash due from related non-banks.[29]

*Non-interest to interest Income*: this variable is the ratio between the total non-interest income and total interest income. The former includes the sum of income from fiduciary activities, service charges on deposit accounts in domestic offices, and trading gains (losses) and fees from foreign exchange transactions, among others. The later includes interest and fee income on loans secured by real estate in domestic offices, interest and fee income on loans to depository institutions in domestic offices, credit cards and related plans, interest income from assets held in trading accounts, among others.

*Non-performing loans:* this variable is the sum of total loans, leasing financing receivables, debt securities and other assets past due 90 days or more.

---

[27] The total fair values are reported as an absolute value.
[28] Banks directly or indirectly owned by the top-tier parent bank holding company, excluding those directly or indirectly owned by the reporting lower-tier parent bank holding company.
[29] Nonbank companies directly or indirectly owned by the top-tier parent bank holding company, excluding those directly or indirectly owned by the reporting lower-tier parent bank holding company.

*Total deposits*: this variable includes the amount of all noninterest-bearing deposits plus the time certificates of deposits of $100,000 or more held in foreign offices of the reporting bank.

*Total loans*: this variable includes all loans except to the commercial paper and the loans reported in the *loan to banks* variable.

# Appendix B

This appendix contains the details on the estimation of the five systemic measures that we consider in this paper. The systemic risk measures are estimated on a weekly basis. In order to conduct quarterly regression analysis we consider the last observation of the quarter. However, for the baseline measure we also consider the sum of the observations during the corresponding quarter as a robustness test.

## B.1. Co-Risk Measures

Adrian and Brunnermeier (2011) based their analysis on the growth rate of the market value of total financial assets, $X_t^i$, which is defined as the growth rate of the product between the market value of institution *i* and its ratio of total assets to book equity.[30] *VaR* and *CoVaR* are estimated by means of quantile regression (Koenker and Bassett, 1978). The time-variant measures are based on the following equations in weekly data:

$$
\begin{aligned}
X_t^i &= \alpha^i + \gamma^i M_{t-1} + \varepsilon_t^i \\
X_t^{system} &= \alpha^{system|i} + \beta^{system|i} X_t^i + \gamma^{system|i} M_{t-1} + \varepsilon_t^{system|i}
\end{aligned} \qquad (B.1.1)
$$

where $M_t^i$ is a set of state variables.[31] In order to perform the quantile regression, we assume a confidence level of 1% what implies to estimate a *VaR* at 1%. Once the coefficients of Equation (B.1.1) have been estimated through quantile regression, we replace them into Equation (B.1.2) to obtain the *VaR* and *CoVaR*.

$$
\begin{aligned}
VaR_t^i(q) &= \hat{\alpha}_q^i + \hat{\gamma}_q^i M_{t-1} \\
CoVaR_t^i(q) &= \hat{\alpha}_q^{system|i} + \hat{\beta}_q^{system|i} VaR_t^i(q) + \gamma_q^{system|i} M_{t-1}
\end{aligned} \qquad (B.1.2)
$$

Finally, the marginal contribution of institution *i* to the overall systemic risk, which is called delta co-value-at-risk ($\Delta CoVaR_i$), is calculated as the difference between $CoVaR_i$ conditional on the distress of the institution (i.e., $q = 0.01$) and the $CoVaR_i$ conditional of the "normal" state of the institution (i.e., $q = 0.5$)

$$\Delta CoVaR_t^i(1\%) = CoVaR_t^i(1\%) - CoVaR_t^i(50\%) \qquad (B.1.3)$$

On the basis of Equation (B.1.3) we obtain the weekly $\Delta CoVaR_t^i$. We also apply this methodology to estimate co-expected shortfall (*CoES_i*) which is defined as the expected shortfall of the financial system conditional on $X^i \leq VaR_q^i$. See Adrian and Brunnermeier (2011) for the details.

---

[30] At portfolio level, the growth rate of the market value of total financial assets is computed as a weighted average of the growth rates of the constituents of the portfolio lagged one period.
[31] This set is composed by VIX, *liquidity spread* (i.e., 3-month repo minus 3-month bill rate), change in 3-month Treasury bill rate, *slope of the yield curve* (i.e., 10-year Treasury rate minus 3-month bill rate), *credit spread* (i.e., 10 Year BAA rated bonds minus 10-year Treasury rate) and return of the MSCI index.

## B.2. Asymmetric CoVaR

Lopez, et al. (2011) propose to extend the $\Delta CoVaR_t^i$ methodology in order to capture asymmetries in the estimation of the co-value-at risk. They propose the following specification:

$$X_t^i = \alpha^i + \gamma^i M_{t-1} + \varepsilon_t^i$$
$$X_t^{system} = \alpha^{system|i} + \beta^{+system|i} X_t^i I_{(X_t^i \geq 0)} + \beta^{-system|i} X_t^i I_{(X_t^i < 0)} + \gamma^{system|i} M_{t-1} + \varepsilon_t^{system|i}$$

$$(B.2.1)$$

where $I_{(\cdot)}$ is an indicator function that takes 1 if the condition of the subscript is true and zero otherwise. Under this specification, Adrian and Brunnermeier (2011) approach can be seen as an special case in which $\beta^{+system|i} = \beta^{-system|i} = \beta^{system|i}$. As in Adrian and Brunnermeier (2011), Equation (B.2.1) is estimated using quantile regression at 1%. Then, $CoVaR_t^i$ is estimated according to Equation (B.2.2):

$$VaR_t^i(q) = \hat{\alpha}_q^i + \hat{\gamma}_q^i M_{t-1}$$
$$CoVaR_t^i(q) = \hat{\alpha}_q^{system|i} + \hat{\beta}_q^{-system|i} VaR_t^i(q) + \gamma_q^{system|i} M_{t-1}$$

$$(B.2.2)$$

## B.3. Gross Shapley Value of Value-at-Risk

In order to apply this methodology it is *sufficient* to define a "characteristic function" ($\vartheta$) which should define the system-wide VaR when it is applied to the entire system. Once the characteristic function have been defined, the contribution of bank $i$ to the subsystem $S$ equals the difference between the risk of subsystem $S$ and the risk of the subsystem when bank $i$ is excluded from it ($S - \{i\}$). So, the Gross Shapley Value ($GSV_i$) equals to the expected value of such contribution when the $N!$ possible orderings may occur with the same probability. Mathematically $GSV_i$ is defined as,

$$GSV_i = \frac{1}{N} \sum_{n_s=1}^{N} \left[ \frac{1}{c(n_s)} \sum_{\substack{S \supset i \\ |S| = n_s}} (\vartheta(S) - \vartheta(S - \{i\})) \right] \quad (B.3.1)$$

where $\Sigma$ denotes the entire financial system, $S \supset i$ are all the possible subsystems in $\Sigma$ containing $i$, $|S|$ represents the number of institutions in the subsystem and $c(n_s)$ comprises the number of all possible subsystem with $n_s$ institutions which is defined as $c(n_s) = \frac{(N-1)!}{(N-n_s)!(n_s-1)!}$.

In order to carry out the practical implementation of this methodology, we estimate the characteristic function as in Adrian and Brunnermeier (2011) (i.e., through quantile regression). The number of considered banks in the system implies the main challenge of this methodology. In this article we analyze 95 bank holding companies and hence, we would have to estimate 2.48E27 different subsystems. Given the unfeasibility of storing such amount of information we define a subset of the 15 largest banks in such a way that for studying every institution we consider 16 banks (i.e., the largest 15 banks plus the bank under study).[32] This modification enables us to

---

[32] The selected banks are: Bank of America, Bank of New York Company, Bank of New York Mellon, BB&T, Charles Schwab, Citigroup, Fifth Third Bancorp, JP Morgan Chase and Company, Metlife, PNC Financial Services Group, State Street, Suntrust Banks, United States Bancorp, Wachovia Corporation and

reduce the size of our problem without biasing the results because those banks represent more than the 80% of the average total assets of the whole system.

Additionally we estimate this measure in an alternative way in which the system (16 banks) is composed of the largest 14 banks, the bank under study and a "synthetic" bank created from the remaining 76 banks which are weighed by the market value of total financial assets. By creating this representative bank, we take all the available information of the system (including the information contained in the small banks). This approach will be considered as a robustness test.

## B.4. Net Shapley Value of Value-at-Risk

We now extend the expression for the GSV for a given bank *i* as presented in Equation (B.3.1) to show that during non-stress periods the individual contribution of this bank to the aggregate systemic risk should be close to zero and consequently this measure will be governed by the individual VaR of bank *i*. To show this, we consider an economy that is composed by 4 banks ($n = 1, ... , 4$). The possible subsystems and the GSV when we study the contribution of bank 1 to the risk of the economy would be:

Subsystems (S): {1}, {1,2}, {1,3}, {1,4}, {1,2,3}, {1,2,4}, {1,3,4}, {1,2,3,4}

$$GSV_1 = \frac{1}{4}\Big[VaR(\{1\}) + \frac{1}{3}$$
$$* \big((VaR(\{1,2\}) - VaR(\{2\})) + (VaR(\{1,3\}) - VaR(\{3\}))$$
$$+ (VaR(\{1,4\}) - VaR(\{4\}))\big) + \frac{1}{3}$$
$$* \big((VaR(\{1,2,3\}) - VaR(\{2,3\})) + (VaR(\{1,2,4\}) - VaR(\{2,4\}))$$
$$+ (VaR(\{1,3,4\}) - VaR(\{3,4\}))\big)$$
$$+ (VaR(\{1,2,3,4\}) - VaR(\{2,3,4\}))\Big] \qquad (B.4.1)$$

In non-stress periods (no systemic risk) bank *i* does not contribute to the overall level of risk and the only term which would differ from zero would be VaR({1}). To check the extent of this problem we estimate the average correlation between the GSV and the VaR for each of the 95 banks. The average correlation for the period 2002-20011 is 0.98. This suggests that GSV is not an appropriate measure in our sample due to their strong correlation with the bank's VaR.

In order to palliate this GSV's drawback we introduce an alternative measure which is free from the impact of the individual value-at-risk. The main reason justifying this adjustment being the VaR$_i$ measures bank *i* specific market risk. But VaR$_i$ does not measure how much risk bank *i* is adding to the whole system. This new measure is named as the Net Shapley Value (NSV$_i$). Mathematically, it is defined as:

$$NSV_i = GSV_i - \frac{1}{N}VaR_i \qquad (B.4.2)$$

Additionally, we estimate the NSV measure for a portfolio that consists of only the 16 largest banks. Note that considering 16 banks we can define the system on the basis of a whole portfolio

---

Wells Fargo and Company. When the bank under study is one of the 15 listed banks, we include the 16th largest bank (Capital One Financial).

of banks instead of focusing on a core subset of banks and adding individually the remaining smaller banks. The pairwise correlation between the NSV as estimated for the baseline analysis and the NSV using a portfolio of the largest 16 banks is, on average, 0.99.

# Appendix C

In this appendix we describe the methodology employed to compare the systemic risk measures described in Appendix B. As in Rodriguez-Moreno and Peña (2013) we use two criteria to compare the five individual contribution of bank to systemic risk measures: (i) the correlation with an index of systemic events and policy actions, and (ii) the Granger causality test.

To implement the first criterion we carry out a regression for each bank *j* in sample, where the dependent variable is the influential event variable (IEV) and the explanatory variable is the systemic risk measure. When the IEV is a categorical variable that takes value 1 whenever there is a systemic event, -1 whenever there is a political action, and 0 otherwise; we run a multinomial logistic regression. When the IEV is a categorical variable that takes value 1 whenever there is a systemic event and 0 otherwise; we run a logistic regression.

$$IEV_t = \alpha + \beta SystemicRiskMeasure_{i,j,t-k} + \varepsilon_t \qquad (C.1)$$

The subindex $i$ refers to a given systemic risk measure (i.e., NSV, GSV, $\Delta CoVaR$, $\Delta CoES$ or asymmetric $\Delta CoVaR$), $j$ refers to bank under analysis ($j = 1, \dots, 95$) and $k$ refers to the number of lags in the regression ($k = 0,1,2$).[33] Next, the McFadden R-squared for each regression is obtained as follows:

$$R^2 = 1 - \frac{ln\hat{L}(M_{Full})}{ln\hat{L}(M_{Intercept})} \qquad (C.2)$$

where $M_{Full}$ refers to the full model and $M_{Intercept}$ to the model without predictors, and $\hat{L}$ is the estimated likelihood.[34] The second criterion is based on the Granger causality test (Granger, 1969). This test examines whether past changes in one variable, $X_t$, help to explain contemporary changes in another variable, $Y_t$. If not, we conclude that $X_t$ does not Granger cause $Y_t$. Formally, the Granger causality test is based on the following regression:

$$\Delta Y_t = \alpha + \sum_{i=1}^{p} \beta_{yi} \Delta Y_{t-i} + \sum_{i=1}^{p} \beta_{xi} \Delta X_{t-i} + \varepsilon_t \qquad (C.3)$$

where $\Delta$ is the first-difference operator and $\Delta X$ and $\Delta Y$ are stationary variables. We reject the null hypothesis that $X_t$ does not Granger cause $Y_t$ if the coefficients $\beta_{xi}$ are jointly significant based on the standard F-test.

---

[33] Results do not change when other lags are considered.
[34] To evaluate the goodness-of-fit for a regression, several pseudo R-squared has been developed. We employ McFadden R-squared due to its appropriate statistical properties.

Table I: Descriptive Statistics of Bank Holding Companies

This table reports the name of the 95 banks that form the sample and related information about their size (average market value in millions of U.S. dollars during the sample period spanning from 1Q2002 to 2Q2011).

| id | Bank Holding | Market Value | id | Bank Holding | Market Value |
|---|---|---|---|---|---|
| 1 | Alabama National Bancorp | 1,063 | 49 | MB Financial | 804 |
| 2 | Amcore Financial | 467 | 50 | Mellon Financial | 16,300 |
| 3 | Associated Banc-Corporation | 2,939 | 51 | Metlife | 31,400 |
| 4 | Bancorpsouth | 1,636 | 52 | National Penn Bancshares | 758 |
| 5 | Bank of America | 140,000 | 53 | NBT Bancorp | 661 |
| 6 | Bank of Hawaii | 2,201 | 54 | New York Community Bancorp | 4,612 |
| 7 | Bank of New York Co | 27,000 | 55 | Newalliance Bancshares | 1,492 |
| 8 | Bank of New York Mellon | 38,100 | 56 | Northern Trust | 12,300 |
| 9 | BB&T | 18,200 | 57 | Old National Bancorp | 1,318 |
| 10 | Bok Financial | 2,589 | 58 | Pacific Capital Bancorp | 941 |
| 11 | Boston Private Financial | 569 | 59 | Park National | 1,230 |
| 12 | Capital One Financial | 16,900 | 60 | PNC Financial Services | 19,600 |
| 13 | Cathay General Bancorp | 1,095 | 61 | Privatebancorp | 588 |
| 14 | Central Pacific Financial | 510 | 62 | Prosperity Bancs | 901 |
| 15 | Charles Schwab | 21,500 | 63 | Provident Bankshares | 644 |
| 16 | Chittenden Corp | 1,119 | 64 | Provident Financial Services | 1,037 |
| 17 | Citigroup | 188,000 | 65 | Regions Financial New | 9,923 |
| 18 | Citizens Republic Bancorp | 970 | 66 | Sky Financial Group | 2,583 |
| 19 | City National | 2,681 | 67 | South Financial Group | 1,012 |
| 20 | Colonial Bancgroup | 1,758 | 68 | State Street | 19,000 |
| 21 | Comerica | 7,893 | 69 | Sterling Bancshares | 621 |
| 22 | Commerce Bancshares | 2,989 | 70 | Sterling Financial | 572 |
| 23 | Community Bank System | 571 | 71 | Suntrust Banks | 18,700 |
| 24 | Cullen Frost Bankers | 2,537 | 72 | Susquehanna Bancshares | 1,004 |
| 25 | CVB Financial | 878 | 73 | SVB Financial Group | 1,503 |
| 26 | East West Bancorp | 1,418 | 74 | Synovus Financial | 6,150 |
| 27 | FNB | 978 | 75 | TCF Financial | 2,986 |
| 28 | Fifth Third Bancorp | 21,300 | 76 | Texas Capital Bancshares | 547 |
| 29 | First Citizens Bancorporation | 411 | 77 | Texas Regional | 1,510 |
| 30 | First Commonwealth Financial | 761 | 78 | Trustmark | 1,488 |
| 31 | First Horizon National | 3,939 | 79 | United States Bancorp | 46,700 |
| 32 | First Midwest Bancorp | 1,280 | 80 | Ucbh Holdings | 921 |
| 33 | First National of Nebraska | 1,222 | 81 | UMB Financial | 1,310 |
| 34 | Firstmerit | 1,935 | 82 | Umpqua Holdings | 817 |
| 35 | Fulton Financial | 2,066 | 83 | United Bankshares | 1,219 |
| 36 | Glacier Bancorp | 765 | 84 | United Community Banks | 721 |
| 37 | Greater Bay Bancorp | 1,315 | 85 | Valley National Bancorp | 2,390 |
| 38 | Hancock Holding | 1,040 | 86 | Wachovia Corp | 48,200 |
| 39 | Harleysville National Corp | 450 | 87 | Webster Financial | 1,762 |
| 40 | Huntington Bancshares | 4,518 | 88 | Wells Fargo and Company | 104,000 |
| 41 | Iberiabank | 583 | 89 | Wesbanco | 530 |
| 42 | International Bancshares | 1,405 | 90 | Westamerica Banc | 1,488 |
| 43 | Investors Bancorp | 1,480 | 91 | Western Alliance Bancorp | 580 |
| 44 | Investors Financial Services | 3,005 | 92 | Whitney Holding Corp | 1,411 |
| 45 | JP Morgan Chase and Co | 117,000 | 93 | Wilmington Trust | 1,924 |
| 46 | Keycorp | 10,200 | 94 | Wintrust Financial | 776 |
| 47 | M&T Bank | 9,396 | 95 | Zions Bancorporation | 5,051 |
| 48 | Marshall & Ilsley | 6,824 | | | |

Table II: Descriptive Statistics

This table reports the descriptive statistics (mean, median, standard deviation, maximum, minimum, and number of observations) of the five groups of determinants of systemic risk under analysis: *size* (log market value); *interconnectedness and substitutability* (commercial paper, loan to banks, total loans, non-interest to interest income, correlation with S&P500, net balances due to banks, net balances due to non-banks); *balance sheet* (leverage, maturity mismatch, total deposits and non-performing loans); *aggregate systemic risk; banks holdings of derivatives* (fair value of credit, interest rate, foreign exchange, equity and commodity derivatives). The descriptive statistics are obtained using quarterly information of the 95 banks reported in Table I from 1Q2002 to 2Q2011.

|  | Mean | Median | Stard. Dev. | Max. | Min. | N. Obs. |
|---|---|---|---|---|---|---|
| *Log market value* | 14.74 | 14.841 | 0.377 | 19.43 | 9.834 | 3077 |
| *Comercial paper/TA* | 0.002 | 0.002 | 0.002 | 0.095 | 0.000 | 3077 |
| *Loan to banks/TA* | 0.002 | 0.002 | 0.002 | 0.071 | 0.000 | 3076 |
| *Total loans/TA* | 0.610 | 0.614 | 0.042 | 0.927 | 0.012 | 3076 |
| *Non-interest to interest income* | 0.484 | 0.478 | 0.119 | 5.305 | -0.648 | 3077 |
| *Correlation with S&P500* | 0.594 | 0.617 | 0.147 | 0.914 | -0.358 | 3077 |
| *Net balance to bank/TA* | 0.000 | 0.000 | 0.000 | 0.019 | -0.023 | 3074 |
| *Net balance to non-bank/TA* | 0.012 | 0.012 | 0.004 | 0.060 | 0.000 | 3074 |
| *Leverage* | 9.540 | 6.620 | 6.881 | 348.7 | 2.714 | 3077 |
| *Maturity mismatch* | 0.094 | 0.094 | 0.034 | 0.640 | 0.000 | 3076 |
| *Total deposits/TA* | 0.687 | 0.689 | 0.038 | 0.905 | 0.001 | 3077 |
| *Non-performing loans/Total loans* | 0.015 | 0.009 | 0.014 | 0.162 | 0.000 | 3077 |
| *Aggregate systemic risk measure* | 0.100 | 0.049 | 0.106 | 0.499 | 0.013 | 3077 |
| *Credit derivatives/TA* | 0.003 | 0.001 | 0.003 | 0.486 | 0.000 | 3074 |
| *Interest rate derivatives/TA* | 0.030 | 0.026 | 0.014 | 1.653 | 0.000 | 3076 |
| *Foreign exchange derivatives/TA* | 0.006 | 0.006 | 0.002 | 0.257 | 0.000 | 3076 |
| *Equity derivatives/TA* | 0.002 | 0.002 | 0.001 | 0.087 | 0.000 | 3074 |
| *Commodity derivatives/TA* | 0.001 | 0.001 | 0.001 | 0.206 | 0.000 | 3074 |

Table III: Correlation Matrix

This table reports the correlation matrix for the variables of interest: size (log market value); interconnectedness and substitutability (commercial paper, loan to banks, total loans, non-interest to interest income, correlation with S&P500, net balances due to banks, net balances due to non-banks); balance sheet (leverage, maturity mismatch, total deposits and non-performing loans); aggregate systemic risk; and banks holdings of derivatives (off-balance sheet derivatives, fair value of credit, interest rate, foreign exchange, equity and commodity derivatives and other derivatives category). The correlations are obtained across the 95 banks reported in Table I for the period spanning from 1Q2002 to 2Q2011.

|  |  | [1] | [2] | [3] | [4] | [5] | [6] | [7] | [8] | [9] | [10] | [11] | [12] | [13] | [14] | [15] | [16] | [17] | [18] |
|---|---|---|---|---|---|---|---|---|---|---|---|---|---|---|---|---|---|---|---|
| [1] | Log market value | 1.000 | | | | | | | | | | | | | | | | | |
| [2] | Commercial paper /TA | 0.529 | 1.000 | | | | | | | | | | | | | | | | |
| [3] | Loan to banks /TA | 0.396 | 0.184 | 1.000 | | | | | | | | | | | | | | | |
| [4] | Total loans /TA | -0.360 | -0.270 | -0.290 | 1.000 | | | | | | | | | | | | | | |
| [5] | Non-interest to interest income | 0.457 | 0.180 | 0.293 | -0.615 | 1.000 | | | | | | | | | | | | | |
| [6] | Correlation with S&P500 | 0.307 | 0.150 | 0.115 | -0.091 | 0.120 | 1.000 | | | | | | | | | | | | |
| [7] | Net balance to bank /TA | 0.072 | 0.053 | 0.038 | 0.009 | 0.036 | 0.011 | 1.000 | | | | | | | | | | | |
| [8] | Net balance to non-bank /TA | -0.114 | 0.014 | 0.049 | 0.097 | -0.052 | 0.036 | 0.074 | 1.000 | | | | | | | | | | |
| [9] | Leverage | -0.226 | -0.007 | -0.026 | 0.052 | -0.055 | -0.149 | 0.078 | 0.091 | 1.000 | | | | | | | | | |
| [10] | Maturity mismatch | 0.229 | 0.218 | 0.052 | -0.190 | 0.046 | 0.115 | 0.045 | -0.029 | 0.006 | 1.000 | | | | | | | | |
| [11] | Total deposits /TA | -0.566 | -0.324 | -0.153 | 0.506 | -0.456 | -0.224 | -0.030 | 0.062 | -0.004 | -0.310 | 1.000 | | | | | | | |
| [12] | Non-performing loans /Total loans | -0.057 | 0.062 | 0.030 | 0.094 | -0.043 | -0.052 | -0.041 | 0.084 | 0.538 | 0.015 | 0.032 | 1.000 | | | | | | |
| [13] | Aggregate systemic risk measue | -0.121 | -0.002 | -0.022 | 0.061 | -0.029 | 0.156 | -0.010 | 0.045 | 0.290 | 0.008 | 0.010 | 0.507 | 1.000 | | | | | |
| [14] | Credit derivatives /TA | 0.292 | 0.270 | 0.158 | -0.183 | 0.029 | 0.091 | -0.027 | 0.061 | 0.095 | 0.106 | -0.230 | 0.164 | 0.128 | 1.000 | | | | |
| [15] | Interest rate derivatives /TA | 0.448 | 0.416 | 0.254 | -0.275 | 0.104 | 0.141 | -0.038 | 0.042 | 0.065 | 0.172 | -0.326 | 0.180 | 0.071 | 0.711 | 1.000 | | | |
| [16] | Foreign exchange derivatives /TA | 0.502 | 0.519 | 0.288 | -0.516 | 0.360 | 0.167 | -0.009 | 0.033 | 0.035 | 0.217 | -0.344 | 0.087 | 0.052 | 0.593 | 0.759 | 1.000 | | |
| [17] | Equity derivatives /TA | 0.468 | 0.446 | 0.245 | -0.308 | 0.101 | 0.145 | -0.035 | 0.031 | 0.052 | 0.159 | -0.346 | 0.114 | 0.014 | 0.676 | 0.851 | 0.757 | 1.000 | |
| [18] | Commodity derivatives /TA | 0.231 | 0.181 | 0.110 | -0.146 | 0.023 | 0.014 | -0.020 | -0.057 | 0.025 | 0.113 | -0.171 | 0.064 | 0.031 | 0.439 | 0.413 | 0.346 | 0.421 | 1.000 |

Table IV: Systemic Risk Measures: Descriptive Statistics and Ranking

This table reports the main descriptive statistics of the systemic risk measures that are individually estimated for the 95 banks reported in Table I and their ranking based on the average McFadden R-squared and Granger causality test. Panel A reports the descriptive statistics of five systemic risk measures in basis points: Net Shapley value (NSV), Gross Shapley Value (GSV), Co-risk measures (ΔCoVaR and ΔCoES), and asymmetric ΔCoVaR. They are reported on quarterly basis calculated at the last week of the corresponding quarter and span from 1Q2002 to 2Q2011. Panel B reports the ranking scores for the systemic risk measures when the influential event variable in Equation (C.1) takes value 1 for those events that should increase systemic risk; -1 whenever there is a policy action that should contribute to decrease systemic risk; and 0 otherwise. Panel C reports the ranking scores for the systemic risk measures for the case in which the influential event variable in Equation (C.1) ignores the policy actions and so, it assigns value 1 whenever there is a systemic event and 0 otherwise. The comparison of different pairs of systemic risk measures, referred to the same bank, based on the McFadden R-squared criterion is done by assigning a score of +1 to the measure with the highest R-squared and -1 to the lowest. The comparison based on the Granger causality test is done by applying the test to pairs of systemic risk measures, referred to the same bank, and giving a score of +1 to measure X if X Granger causes another measure Y at 5% confidence level and -1 if X is caused in the Granger sense by Y. Finally we add up the scores obtained by each measure across the 95 banks to obtain the one with highest score.

| Panel A | | | | | | |
|---|---|---|---|---|---|---|
| | Mean | Median | Stard. Dev. | Max. | Min. | N. Obs. |
| *Net Shapley Value* | 11.08 | 6.41 | 11.27 | 176.39 | -76.03 | 3077 |
| *Gross Shapley Value* | 92.61 | 82.09 | 48.59 | 546.15 | 8.99 | 3077 |
| *Delta co-value-at-risk* | 752.02 | 641.05 | 493.47 | 3201.14 | 19.05 | 3077 |
| *Delta co expected shortfall* | 458.67 | 398.48 | 310.87 | 2215.91 | -307.62 | 3077 |
| *Asymmetric Delta co-value-at-risk* | 769.11 | 658.62 | 496.05 | 4327.27 | -151.70 | 3077 |

| Panel B | | | | | |
|---|---|---|---|---|---|
| | Net Shapley Value | Gross Shapley Value | Delta co-value-at-risk | Delta co-expected-shortfall | Asymmetric Delta co-value-at-risk |
| *McFadden R-squared* | 246 | 68 | -20 | -272 | -22 |
| *Granger causality test* | 11 | 8 | -18 | -1 | 0 |
| *Total* | 257 | 76 | -38 | -273 | -22 |
| *Average McFadden R-squared* | 0.2045 | 0.1544 | 0.1361 | 0.1127 | 0.1374 |

| Panel C | | | | | |
|---|---|---|---|---|---|
| | Net Shapley Value | Gross Shapley Value | Delta co-value-at-risk | Delta co-expected-shortfall | Asymmetric Delta co-value-at-risk |
| *McFadden R-squared* | 272 | -8 | 4 | -256 | -12 |
| *Granger causality test* | 11 | 8 | -18 | -1 | 0 |
| *Total* | 283 | 0 | -14 | -257 | -12 |
| *Average McFadden R-squared* | 0.1188 | 0.0869 | 0.0817 | 0.0721 | 0.0829 |

## Table V: Baseline Regression

This table reports the effects of the determinants of systemic risk that is measured as the Net Shapley Value (in basis points). Our sample is composed of the 95 banks reported in Table I and spans from 1Q2002 to 2Q2011. Column 1 reports the results of Equation (1) whose coefficients are estimated by means of a Prais-Winsten regression robust to heteroskedasticity, contemporaneous correlation across panels. Column 2 contains the economic impact of the statistically significant variables (in percentage) and it is reported for the statistically significant coefficients at 1 or 5% levels. The economic impact is defined as the ratio of the coefficients in Column 1 times the standard deviation of the corresponding explanatory variable over the mean of the dependent variable. Column 3 reports the estimated coefficients for Equation (1) when the holdings of derivatives are not used in the estimation. Column 4 reports the estimated coefficients on the basis of a fixed-effect regression at bank level with Driscoll and Kraay (1998) standard errors which are robust to heteroskedasticity, autocorrelation and correlated panels. The symbol *** (**) denotes the significance level at 1% (5%). The results correspond to the estimated coefficient and the robust standard errors (between brackets).

|  | (1) Coefficient [SE] | (2) Economic Impact (%) | (3) Coefficient [SE] | (4) Coefficient [SE] |
|---|---|---|---|---|
| Log market value $_{t-1}$ | -1.939 [3.070] |  | -1.139 [2.908] | -0.825 [7.772] |
| Log of squared market value $_{t-1}$ | 0.022 [0.099] |  | -0.002 [0.093] | -0.091 [0.280] |
| Commercial paper $_{t-1}$ /TA | 32.492 [31.301] |  | 76.700** [30.500] | 47.380 [37.110] |
| Loan to banks $_{t-1}$ /TA | 5.698 [42.920] |  | -12.720 [41.440] | -176.700 [87.120] |
| Total loans $_{t-1}$ /TA | 8.684*** [2.760] | 3.221 | 5.056** [2.420] | 17.470 [13.020] |
| Non-interest to interest income $_{t-1}$ | 0.348 [0.857] |  | 1.072 [0.747] | -3.613 [2.490] |
| Correlation with S&P500 $_{t-1}$ | 2.291 [2.883] |  | 2.454 [2.904] | 1.184 [3.059] |
| Net balance to bank $_{t-1}$ /TA | 439.316*** [96.461] | 1.576 | 437.600*** [97.260] | 604.4*** [189.0] |
| Net balance to non-bank $_{t-1}$ /TA | 8.005 [26.636] |  | 20.610 [26.740] | 30.170 [30.500] |
| Leverage $_{t-1}$ | 0.152*** [0.039] | 9.285 | 0.162*** [0.0404] | 0.061** [0.028] |
| Maturity mismatch $_{t-1}$ | -0.993 [2.714] |  | 0.380 [2.889] | -6.716 [3.896] |
| Total deposits $_{t-1}$ /TA | -17.720*** [3.398] | -6.034 | -14.550*** [3.439] | -35.850** [15.62] |
| Non-performing loans $_{t-1}$ /Total loans | 157.471*** [48.660] | 19.056 | 148.800*** [49.700] | 162.000*** [50.590] |
| Aggregate systemic risk measue $_{t-1}$ | 70.520*** [16.193] | 66.485 | 72.290*** [16.550] | 66.560*** [3.445] |
| Aggregate systemic risk measue $_{t-2}$ | -24.544 [16.283] |  | -25.080 [16.610] | -21.560** [10.090] |
| Credit derivatives $_{t-1}$ /TA | 35.285*** [8.495] | 0.968 |  | 32.430** [12.450] |
| Interest rate derivatives $_{t-1}$ /TA | -12.947*** [3.291] | -1.608 |  | -15.550** [7.435] |
| Foreign exchange derivatives $_{t-1}$ /TA | 94.358*** [24.846] | 1.918 |  | 129.800** [57.310] |
| Equity derivatives $_{t-1}$ /TA | -29.785 [45.249] |  |  | 79.830 [73.950] |
| Commodity derivatives $_{t-1}$ /TA | -19.698 [12.391] | -0.196 |  | 8.689 [10.020] |
| Constant | 31.605 [24.218] |  | 24.460 [22.980] | 50.610 [54.52] |
| Number of Observations | 3070 |  | 3070 | 3070 |
| Number of Groups | 95 |  | 95 | 95 |
| Min. Observations per Group | 13 |  | 13 | 13 |
| Avg. Observations per Group | 32.32 |  | 32.32 | 32.32 |
| Max. Observations per Group | 36 |  | 36 | 36 |
| R-squared | 0.486 |  | 0.477 | 0.4811 |

## Table VI: Two-Stage Selection Model

This table reports the results of a Heckman two-stage selection model that is applied to a sample of 95 banks that spans from 1Q2002 to 2Q2011 to analyze the determinants of systemic risk. In the first stage (Panel A) we estimate a Probit regression model in which the dependent variable is a dummy that indicates whether the bank uses or not each type of derivative (one column for each derivative) and the explanatory variables refer to bank-specific information and the holdings of the remaining four types of derivatives. The Probit model also includes time effects and the estimation is robust to heteroskedatiscity. From the first-stage selection model we get the inverse Mills ratio that will be used as an input in the second-stage regression. In the second-stage (Panel B) we estimate the baseline model by means of a Prais-Winsten robust to heteroskedasticity and contemporaneous correlation across panels. In this estimation, the dependent variable is the individual contribution to systemic risk measured from the Net Shapley Value (in basis points). Panel B also consists of five columns, one for each type of derivative. The explanatory variables in each column include the self-selection parameter for each derivative and the remaining variables employed in the baseline analysis. We exclude from the regression the four types of derivatives that were used in the first-stage to compute the total holdings of the remaining types of derivatives. The symbol *** (**) denotes the significance level at 1% (5%). The results correspond to the estimated coefficient and the robust standard errors (between brackets).

|  | (1) Credit Deriv. Coefficient [SE] | (2) FE Deriv. Coefficient [SE] | (3) IR Deriv. Coefficient [SE] | (4) Equity Deriv. Coefficient [SE] | (5) Commodity Deriv. Coefficient [SE] |
|---|---|---|---|---|---|
| Leverage $_{t-1}$ | 0.013*** [0.002] | 0.009*** [0.002] | 0.131*** [0.031] | 0.006*** [0.002] | 0.008*** [0.002] |
| Non-performing loans $_t$ /Total loans | -0.389 [3.139] | 5.406** [2.558] | 16.813*** [4.876] | 0.259 [2.507] | 7.207*** [2.164] |
| Log market value $_t$ | 0.812*** [0.045] | 0.604*** [0.047] | 0.352*** [0.049] | 0.508*** [0.031] | 0.552*** [0.032] |
| Correlation with S&P500 $_t$ | 0.739** [0.373] | 0.057 [0.192] | 0.907*** [0.205] | -0.013 [0.222] | -0.031 [0.255] |
| Non-interest to interest income $_t$ | 0.498*** [0.091] | -0.130 [0.105] | 2.077*** [0.258] | 0.160*** [0.059] | 0.046 [0.065] |
| Commercial paper $_t$ /TA | 10.849 [7.225] | 538.622*** [84.849] |  | 32.388*** [6.774] | -14.796*** [5.676] |
| Total loans $_t$ | 1.800*** [0.343] |  | 1.712*** [0.372] |  |  |
| Loans to foreign banks and goverments $_t$ |  | 0.732*** [0.102] |  |  |  |
| Loans to agricultural production $_t$ |  |  |  |  | 13.985*** [2.445] |
| Deriv. other than credit derivatives $_t$ | 15.562*** [2.377] |  |  |  |  |
| Deriv. other than FE derivatives $_t$ |  | 90.629*** [10.479] |  |  |  |
| Deriv. other than IR derivatives $_t$ |  |  | 62.052*** [18.872] |  |  |
| Deriv. other than equity derivatives $_t$ |  |  |  | 7.583*** [1.624] |  |
| Deriv. other than commodity derivatives $_t$ |  |  |  |  | 2.930*** [0.490] |
| Constant | -17.413*** [0.706] | -9.316*** [0.666] | -7.290*** [0.834] | -8.774*** [0.516] | -9.792*** [0.513] |
|  | 3,074 | 3,071 | 3,071 | 3,071 | 3,071 |

Table VI (con't): Two-Stage Selection Model

This table contains the second-stage of the Heckman two-stage selection model.

|  | (1) Coefficient [SE] | (2) Coefficient [SE] | (3) Coefficient [SE] | (4) Coefficient [SE] | (5) Coefficient [SE] |
|---|---|---|---|---|---|
| *Log market value $_{t-1}$* | -7.549 [4.245] | -1.826 [4.368] | -2.763 [2.991] | -10.196 [5.544] | -4.682 [3.676] |
| *Log of squared market value $_{t-1}$* | 0.138 [0.111] | -0.003 [0.133] | 0.042 [0.097] | 0.177 [0.129] | 0.046 [0.095] |
| *Commercial paper $_{t-1}$ /TA* | 63.084 [32.677] | 39.194 [35.005] | 79.459** [31.283] | -11.685 [52.495] | 108.861*** [37.613] |
| *Loan to banks $_{t-1}$ /TA* | -23.950 [42.354] | -8.594 [43.716] | -7.575 [43.911] | -47.466 [40.630] | -20.101 [41.373] |
| *Total loans $_{t-1}$ /TA* | 1.691 [3.262] | 7.415*** [2.682] | 2.322 [2.512] | 4.658 [2.472] | 4.820* [2.508] |
| *Non-interest to interest income $_{t-1}$* | 0.179 [1.184] | 0.756 [0.799] | -0.017 [1.056] | -0.262 [1.146] | 0.953 [0.770] |
| *Correlation with S&P500 $_{t-1}$* | 0.876 [3.205] | 2.509 [2.998] | 1.830 [3.091] | 2.466 [2.938] | 2.492 [2.962] |
| *Net balance to bank $_{t-1}$ /TA* | 515.983*** [97.798] | 464.783*** [100.978] | 435.440*** [100.104] | 443.972*** [97.122] | 475.926*** [96.074] |
| *Net balance to non-bank $_{t-1}$ /TA* | -0.987 [19.717] | 16.790 [26.251] | 26.020 [27.076] | 2.875 [22.643] | 13.763 [22.855] |
| *Leverage $_{t-1}$* | 0.114** [0.050] | 0.148*** [0.043] | 0.149*** [0.041] | 0.099** [0.052] | 0.123*** [0.047] |
| *Maturity mismatch $_{t-1}$* | 0.570 [2.879] | -0.678 [2.712] | -1.020 [3.066] | -0.252 [2.959] | 0.489 [2.889] |
| *Total deposits $_{t-1}$ /TA* | -14.501*** [3.485] | -15.936*** [3.432] | -14.454*** [3.518] | -14.947*** [3.605] | -14.546*** [3.531] |
| *Non-performing loans $_{t-1}$ /Total loans* | 133.307*** [48.615] | 144.558*** [51.317] | 133.539*** [50.925] | 141.483*** [49.586] | 125.939** [54.093] |
| *Aggregate systemic risk measue $_{t-1}$* | 67.925*** [16.797] | 71.669*** [16.696] | 72.376*** [16.756] | 68.399*** [16.689] | 70.676*** [16.721] |
| *Aggregate systemic risk measue $_{t-2}$* | -30.415 [16.962] | -25.861 [16.803] | -25.548 [16.810] | -31.193 [16.974] | -26.084 [16.620] |
| *Inverse Mills Ratio $_{t-1}$* | -2.769 [1.738] | -1.174 [0.878] | -3.485** [1.845] | -8.149 [4.567] | -4.342 [3.348] |
| *Credit derivatives $_{t-1}$ /TA* | 31.642*** [10.453] | | | | |
| *Foreign exchange derivatives $_{t-1}$ /TA* | | 49.171*** [16.365] | | | |
| *Interest rate derivatives $_{t-1}$ /TA* | | | -1.307 [1.492] | | |
| *Equity derivatives $_{t-1}$ /TA* | | | | 21.862 [26.968] | |
| *Commodity derivatives $_{t-1}$ /TA* | | | | | -15.644 [16.626] |
| *Constant* | 99.618** [46.474] | 35.819 [36.595] | 42.687 [23.968] | 136.051** [64.430] | 75.878 [42.418] |
| Observations | 2,992 | 2,992 | 2,992 | 2,992 | 2,992 |
| R-squared | 0.487 | 0.482 | 0.481 | 0.486 | 0.483 |
| Number of Groups | 95 | 95 | 95 | 95 | 95 |

Table VII: Alternative analyses of potential endogeneity bias

This table contains alternative analyses of the potential endogeneity bias. The regressions are at bank level (95 banks) and for the period 1Q2002 to 2Q2011. Column 1 contains the results based on the Westerlund and Narayan (2012) two-stage estimation. We first compute the adjusted systemic risk as the residual of regressing systemic risk on all the explanatory variables of the baseline analysis apart from the derivatives variables. We then check the endogeneity of derivatives holdings and for those derivatives found to be endogenous we implement the two-stage analysis. In the first-stage we regress the holdings of an individual derivative on its own lag. In the second-stage we include, besides the holdings of that derivative, the correction term that is the residual obtained in the first-stage. We report the estimation of the second-stage which simultaneously uses all the derivatives with endogeneity problems: credit, foreign exchange, and interest rate derivatives. Column 2 reports the effect of the fair value of the (endogenous) holdings of credit, foreign exchange, and interest rate derivatives on systemic risk when they are instrumented with information on the bank trading activity and its expertise in the three previous derivatives. The first group of instruments contains: assets in trading accounts and the total fair value of derivatives held for trading relative to total assets, both lagged one quarter. The second group of instruments contains: the fair value of the holdings of each derivative relative to the sum of the face value for the five types of derivatives used in this study lagged one year. The regression is performed on the basis of a fixed effect robust to heterostkedasticty regression. At the bottom of Column 2 we report the results for the underidentification and overidentification tests that enable us to evaluate the appropriateness of the instruments. The symbol *** (**) denotes the significance level at 1% (5%). The results correspond to the estimated coefficient and the robust standard errors (between brackets).

| | (1) | (2) |
|---|---|---|
| Credit derivatives $_{t-1}$ /TA | 32.855*** [10.121] | 55.256** [24.017] |
| Interest rate derivatives $_{t-1}$ /TA | -10.72*** [2.138] | -42.841*** [18.930] |
| Foreign exchange derivatives $_{t-1}$ /TA | 48.897*** [11.140] | 343.466*** [140.26] |
| Correction for endogeneity | YES | |
| Control variables | | YES |
| Number of Observations | 3070 | 2352 |
| R-squared | 0.0309 | 0.4586 |
| Underidentification test | Kleibergen-Paap rk LM sta Chi-sq(3) P-val | 24.188 0.000 |
| Overidentification test of all instruments | Hansen J statistic Chi-sq(2) P-val | 0.987 0.611 |
| Instrumented | Credit, interest and foreign derivatives to total assets (lagged 1 quarter) | |
| Excluded instrument | Assets in trading accounts (lagged 1 quarter), face value of derivatives held for trading purposes (lagged 1 quarter), and fair value of credit, foreign exchange, and interest rate derivatives relative to the far value of the total holdings of derivatives (lagged 1 year) | |

Table VIII: Difference-in-difference analysis

This table contains the results for a difference-in-difference analysis in which we consider the bankruptcy filing of Lehman Brothers on September 15, 2011 (2008Q3) as an exogenous shock to deal with potential endogeneity bias. Banks with more holdings of derivatives relative to total assets are defined as the treatment group and banks with fewer holdings are the control group. Banks are ranked according to their average holdings in the second and third quarters of year 2007. The dummy variable of *Top-quartile* is set to unity if the bank's holdings of a given type of derivative is in the top-quartile (75-percentile and above), and zero if it is in the bottom-quartile (25-percentile and below). The dummy variable of *Post-Lehman* is set to unity if the date is 2008Q4 (the quarter after the bankruptcy filing of Lehman Brothers), and zero if the date is 2007Q4 (one year before the bankruptcy filing of Lehman Brothers). A third dummy variable *Top-quartile\*Post-Lehman* is the cross-product of the previous two dummy variables. The diff-in-diff analysis is implemented individually for each derivative with potential problems of endogeneity: credit, foreign exchange, and interest rate derivatives. The control variables employed in the baseline analysis and the derivatives that are found to be exogenous are also used as explanatory variables. The symbol \*\*\* (\*\*) denotes the significance level at 1% (5%). The results correspond to the estimated coefficient and the robust standard errors (between brackets).

|  | Credit Deriv. | FE Deriv. | IR Deriv. |
|---|---|---|---|
| *Post Lehman (Dummy 2008Q4)* | 27.862*** | 28.204*** | 29.223*** |
|  | [1.979] | [2.15] | [2.404] |
| *Top-quartile* | 18.265*** | 10.389*** | 5.996 |
|  | [4.267] | [3.963] | [3.633] |
| *Post Lehman x Top-quartile* | -.108 | -1.426 | -5.415** |
|  | [4.121] | [2.994] | [2.717] |
| *Control variables* | YES | YES | YES |
| Number of Observations | 178 | 178 | 178 |
| R-squared | 0.873 | 0.861 | 0.856 |

Table IX: Alternative Measures of Systemic Risk

This table reports the results of a variation in the baseline unbalanced panel regression (Equation 1) in which different specifications of the dependent variable (contributions to systemic risk) are considered while the explanatory variables employed do not change. In the interest of brevity we only report the effect of derivative holdings. Our database is composed of the 95 banks reported in Table I and spans from 1Q2002 to 2Q2011. We estimate the coefficients by means of a Prais-Winsten robust to heteroskedasticity, contemporaneous correlation across panels. This table reports the results of using alternative contributions to systemic risk: (1) Net Shapley Value at the end of the quarter (baseline); (2) Net Shapley Value using the alternative approach at the end of the quarter; (3) sum of the Net Shapley Value for the corresponding quarter; (4) Gross Shapley Value the end of the quarter; (5) Delta Co-Value-at-Risk (ΔCoVaR); (6) Delta Co-Expected-Shortfall (ΔCoES); and (7) Asymmetric Delta Co-Value-at-Risk (Asymmetric ΔCoVaR). All dependent variables are measured on basis points but due to the different nature of each measure the magnitude of the coefficients is not comparable across the different columns. The results presented correspond to the estimated coefficient and the robust standard errors. The symbol *** (**) denotes that the variable is significant at 1% (5%).

|  | (1) | (2) | (3) | (4) | (5) | (6) | (7) |
|---|---|---|---|---|---|---|---|
|  | *Coefficient* | *Coefficient* | *Coefficient* | *Coefficient* | *Coefficient* | *Coefficient* | *Coefficient* |
|  | *[SE]* | *[SE]* | *[SE]* | *[SE]* | *[SE]* | *[SE]* | *[SE]* |
| *Credit derivatives $_{t-1}$ /TA* | 35.29*** | 34.90*** | 446.4*** | 195.8*** | 733.3** | 697.4** | 1128.9** |
|  | [8.49] | [8.53] | [108.3] | [41.88] | [370.38] | [270.4] | [469.0] |
| *Interest rate derivatives $_{t-1}$ /TA* | -12.95*** | -12.97*** | -156.3*** | -78.88*** | -590.2*** | -398.9*** | -860.1*** |
|  | [3.291] | [3.293] | [41.43] | [16.29] | [134.1] | [84.88] | [139.3] |
| *Foreign exchange derivatives $_{t-1}$ /TA* | 94.36*** | 96.73*** | 1134.6*** | 536.0*** | 4347.9*** | 3099.7*** | 4571.4*** |
|  | [24.85] | [24.93] | [242.6] | [103.3] | [764.9] | [487.5] | [725.0] |
| *Equity derivatives $_{t-1}$ /TA* | -29.78 | -24.56 | -294.1 | -26.40 | -1635.9 | -1911.8 | 1036.7 |
|  | [45.25] | [45.17] | [551.1] | [247.9] | [1897.1] | [1324.7] | [2045.6] |
| *Commodity derivatives $_{t-1}$ /TA* | -19.70 | -18.79 | -283.1 | -228.9*** | 341.8 | -867.1 | -306.2 |
|  | [12.39] | [12.35] | [172.8] | [68.55] | [789.5] | [533.0] | [748.2] |
| *Control Variables* | Yes | Yes | Yes | Yes | Yes | Yes | Yes |
| Number of Observations | 3070 | 3070 | 3070 | 3070 | 3070 | 3070 | 3070 |
| Number of Groups | 95 | 95 | 95 | 95 | 95 | 95 | 95 |
| Min. Observations per Group | 13 | 13 | 13 | 13 | 13 | 13 | 13 |
| Avg. Observations per Group | 32.32 | 32.32 | 32.32 | 32.32 | 32.32 | 32.32 | 32.32 |
| Max. Observations per Group | 36 | 36 | 36 | 36 | 36 | 36 | 36 |
| R-squared | 0.491 | 0.492 | 0.606 | 0.362 | 0.382 | 0.315 | 0.370 |

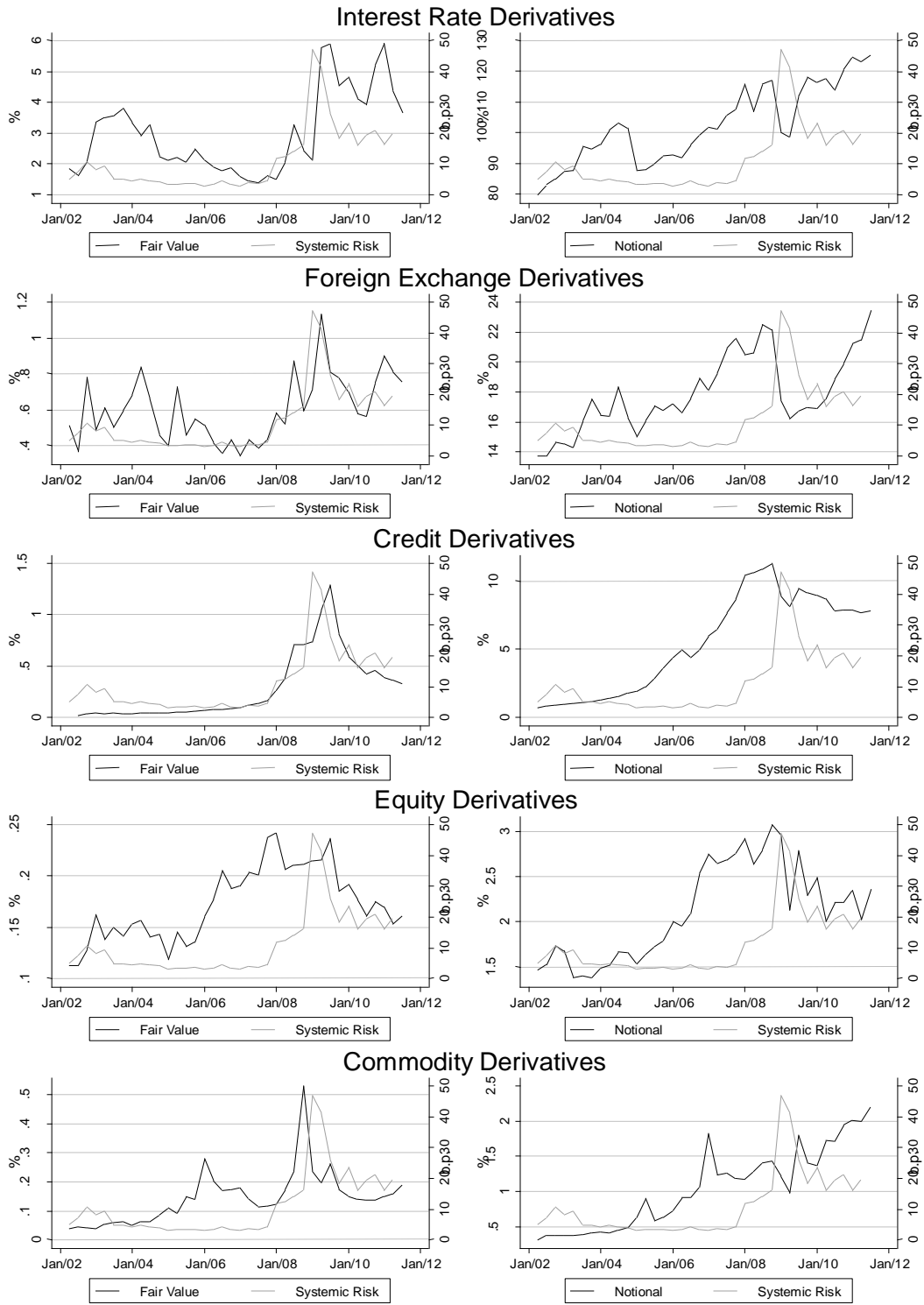

**Figure 1: Systemic risk measure and banks' holdings of derivatives relative to total assets**. Figure 1 depicts the average fair values (on the left-hand-side) and notional amounts (right-hand-side) of the banks holdings of (by order) interest rate, foreign exchange, credit, equity, and commodity derivatives over total assets across the 95 banks in the sample for the period 1Q2002-2Q2011. These ratios are expressed in percentages and are lagged one period *(t-1)* while the average systemic risk measure defined as the average Net Shapley Value, which is also contained in each panel of the figure, is depicted at period *t* such as they appear in Equation (1). Note that the y-axes have different scales for the different derivatives for a better view of their trend.